\documentclass[prb,superscriptaddress,showpacs,floatfix,tightenlines,10pt,twocolumn]{revtex4}
\usepackage{amssymb}
\usepackage{amsmath}
\usepackage{graphicx}

\begin{document}

\title{Bilayer graphene as an helical quantum Hall ferromagnet}
\author{R. C\^{o}t\'{e} }
\affiliation{D\'{e}partement de physique, Universit\'{e} de Sherbrooke, Sherbrooke, Qu%
\'{e}bec, J1K 2R1, Canada}
\author{J. P. Fouquet}
\affiliation{D\'{e}partement de physique, Universit\'{e} de Sherbrooke, Sherbrooke, Qu%
\'{e}bec, J1K 2R1, Canada}
\author{Wenchen Luo}
\affiliation{D\'{e}partement de physique, Universit\'{e} de Sherbrooke, Sherbrooke, Qu%
\'{e}bec, J1K 2R1, Canada}

\begin{abstract}
The two-dimensional electron gas in a bilayer graphene in the Bernal
stacking supports a variety of uniform broken-symmetry ground states in
Landau level $N=0$ at integer filling factors $\nu \in \left[ -3,4\right] .$
When an electric potential difference (or bias) is applied between the
layers at filling factors $\nu =-1,3$, the ground state evolves from an
interlayer coherent state at small bias to a state with orbital coherence at
higher bias where \textit{electric} dipoles associated with the orbital
pseudospins order spontaneously in the plane of the layers. In this paper,
we show that by further increasing the bias at these two filling factors,
the two-dimensional electron gas goes first through a Skyrmion crystal state
and then into an helical state where the pseudospins rotate in space. The
pseudospin textures in both the Skyrmion and helical states are due to the
presence of a Dzyaloshinskii-Moriya interaction in the effective pseudospin
Hamiltonian when orbital coherence is present in the ground state. We study
in detail the electronic structure of the helical and Skyrmion crystal
states as well as their collective excitations and then compute their
electromagnetic absorption.
\end{abstract}

\date{\today }
\maketitle

\section{INTRODUCTION}

Experiments\cite{Novoselov} \ have shown that, when placed in a
perpendicular magnetic field, a graphene bilayer supports a set of Landau
levels with energies given by $E_{N}=\pm \hslash \omega _{c}\sqrt{\left\vert
N\right\vert \left( \left\vert N\right\vert +1\right) }$ where $N=0,\pm
1,\pm 2,\ldots $ All levels except $N=0$ are fourfold degenerate in addition
to the usual degeneracy related to the guiding center coordinate. An
electronic state is thus specified by its Landau level index $N,$ its
guiding center index $X$ (in the Landau gauge), its spin index $\sigma =\pm
1 $ and its valley index $\xi =\pm K.$ In $N=0,$ we must add an additional%
\textit{\ orbital} index because states with both $n=0$ and $n=1$
Landau-level character have zero kinetic energy. The eightfold degeneracy of
the $N=0$ Landau level is revealed experimentally by a jump in the quantized
Hall conductivity from $-4\left( e^{2}/h\right) $ to $+4\left(
e^{2}/h\right) $ when the charge density is tuned across neutrality in
moderately disordered samples\cite{Octetexperiments}.

Some of us\cite{BarlasPRL1} have shown that the close proximity between the
two graphene layers in bilayer graphene leads to spontaneous interlayer
coherence in $N=0$ when Coulomb interaction is taken into account. In a
pseudospin language where the spin, layer and orbital degrees of freedom are
each mapped to a $S_{i}=1/2$ pseudospin ($i=\sigma ,\xi ,n),$
electron-electron interactions at integer filling factors completely lifts
the degeneracy of the bilayer octet producing spontaneously broken-symmetry
ground states with spin, valley and orbital polarizations. In consequence,
quantum Hall plateaus should occur at all integer values of the filling
factors from $\nu =-3$ to $\nu =4$ in Landau level $N=0.$ The existence of
these additional plateaus has recently been confirmed experimentally in
suspended bilayer graphene samples and bilayer graphene on SiO$_{2}$/Si
substrates\cite{Octetexperiments}.

The possibility to study novel broken-symmetry (BS)\ states is always
exciting and the new BS states in bilayer graphene are no exceptions. For
example, the interlayer-coherent state at $\nu =-3$ (and $\nu =1$ if the
two-dimensional electron gas (2DEG) can be considered as fully spin
polarized) is a quantum Hall layer-pseudospin ferromagnet with pseudospin
wave dispersion $\omega \sim q^{2}$ that contrasts with the usual linear
dispersion found in interlayer-coherent states in semiconductor
double-quantum-well systems. This unusual dispersion can be related\cite%
{BarlasPRL2} to a vanishing of the counterflow superfuid density. At filling
factor $\nu =-1,3$, the interlayer-coherent state has the usual linear
dispersion\cite{CoteOrbital}. If a positive electric potential difference $%
\Delta _{B}$ (which we refer to, in this paper, as the bias potential) is
applied between the layers, the charge is progressively transferred into the
bottom layer. At $\nu =-3,1$ the ground state with all the charge in one
layer supports an orbital pseudospin mode that can be viewed as an \textit{%
intralayer} cyclotron resonance\cite{BarlasPRL1}. This mode is gapped due to
the finite bias and should be detectable in microwave absorption experiments.

The broken-symmetry states related to the orbital degree of freedom are
especially interesting. For a spin-polarized 2DEG, they occur at filling
factors $\nu =-1,3$ and above a certain critical bias $\Delta _{B}$ where
one of the layer is completely filled. The homogeneous orbital
broken-symmetry states have a finite density of electric dipoles that
collectively order in the $x-y$ plane\cite{CoteOrbital,Shizuya1}. The
orbital pseudospin mode corresponding to the collective motion of these
dipoles is gapless despite the finite bias. It is a Goldstone mode due to
the breaking of the $U(1)$ symmetry of the pseudospins in the $x-y$ plane.

It was shown recently\cite{CoteOrbital} that there exists a
Dzyaloshinskii-Moriya (DM) interaction between the orbital pseudospins at $%
\nu =-1,3$ that causes the orbital pseudospin mode to soften at a finite
wave vector $\mathbf{q}$ as the bias $\Delta _{B}$ is increased. It was
conjectured that, above a critical bias, the ground state should be some
kind of helical state. In the present work, we find a more complex scenario.
Working in the Hartree-Fock approximation, we find that, as the bias is
increased, the ground state follows the sequence of transitions: uniform
state - Skyrmion crystal - helical state - Skyrmion crystal - uniform state.
The phase diagram is symmetrical about the bias $\Delta _{B}^{(1)}/2$ where
the charge is equally distributed in both orbitals $n=0$ and $n=1.$
Interestingly, our phase diagram looks very similar to that found recently
in a thin film of the helical magnet Fe$_{0.5}$Co$_{0.5}$Si where a Skyrmion
crystal as well as an helical and a ferromagnetic uniform phases have all
been observed using Lorentz transmission electron microscopy\cite{XZYu,jhhan}%
. In this system, the phase transitions are induced by a transverse magnetic
field. In the bilayer, the role of the magnetic field is played by the bias.
The effective pseudospin hamiltonian of the orbital states in a graphene
bilayer is quite complex. Since charge and pseudospin are coupled, a
pseudospin texture such as that found in the orbital Skyrmion crystal
produces a charge density in real space and so the direct (or Hartree)
Coulomb interaction must be considered together with the other competing
interactions mentionned above.

In this work, we study in detail the helical and Skyrmion crystal states. We
derive their electronic properties as well as their collective excitations
and compute their electromagnetic absorption. We show that the effective
pseudospin model that describes these states involves nonlocal as well as
anisotropic exchange interactions between the pseudospins. These Coulomb
exchange interaction tend to align the orbital pseudospins while the DM term
favors a rotation of the pseudospins in space. The resulting ground states
result from a balance between these competing forces just as in helical
magnets such as MnSi and Fe$_{1-x}$Co$_{x}$Si. One major difference in the
graphene bilayer is that the DM interaction in the orbitally coherent state
is of an entirely different origin than the spin-orbit interaction at work
in usual helical magnets\cite{CoteOrbital}.

Our paper is organized in the following way. In Sec. II, we study the
non-interacting states of the graphene bilayer within a two-band low-energy
model. In Sec. III, we discuss the validity of our approximation of assuming
complete spin polarization. This discussion is needed since increasing the
bias pushes down(up) half of the spin down(up) levels. Some levels of
opposite spin cross at finite bias and this can introduce new phases in the
phase diagram or make some phases disappear\cite{Lambert}. In Sec. IV, we
summarize the Hartree-Fock approximation as well as the generalized
random-phase approximation\ (GRPA) which we use to study the collective
excitations. In Sec. V, we study the electronic properties of the different
phases as well as their collective excitations. We discuss their
electromagnetic absorption in Sec. VI and conclude in Sec. VII.

\section{EFFECTIVE\ TWO-BAND MODEL}

We consider a graphene bilayer in Bernal stacking as represented in Fig. \ref%
{figbilayer}. The bilayer is placed in an external transverse \textit{%
electric} field in order to allow an electrical potential difference $\Delta
_{B}$ between the layers and in a transverse \textit{magnetic} field $%
\mathbf{B}.$ We denote the two basis atoms in the top layer by $A_{1}$ and $%
B_{1}$ and those of the bottom layer by $A_{2}$ and $B_{2}$ with atoms $%
A_{1} $ sitting directly above atoms $B_{2}$ as shown in the figure. The
band structure of this system is calculated in the tight-binding
approximation with the hopping parameters: $\gamma _{0}=3.12$ $\mathrm{eV,}%
\gamma _{1}=0.39 $ $\mathrm{eV,}\gamma _{4}=0.12$ $\mathrm{eV,}\Delta
=0.0156 $ $\mathrm{eV,}$ taken from Ref. \onlinecite{Castrojphys}. The
parameter $\gamma _{0}$ is the intralayer hopping term between
nearest-neighbor carbon atoms, $\gamma _{1}$ is the interlayer hopping
between carbon atoms that are part of a dimer, $\gamma _{4}$ is an
interlayer hopping term between two carbons atoms that are not part of a
dimer ($A_{1}-A_{2}$ and $B_{1}-B_{2}$) and $\Delta $ represents the
difference in the crystal field experienced by the inequivalent atoms $A_{i}$
and $B_{i}$ in the same plane. We neglect the trigonal warping term $\gamma
_{3}$ (the $A_{2}-B_{1}$ hopping), a correct approximation at sufficiently
high magnetic field\cite{Mccann}.

\begin{figure}[tbph]
\includegraphics[scale=1]{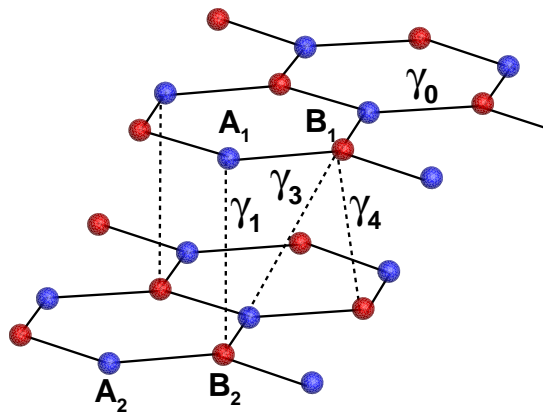}
\caption{(Color online) Lattice structure of the bilayer graphene in the
Bernal stacking. The spheres represent carbon atoms.}
\label{figbilayer}
\end{figure}

The electronic dispersion consists of four bands. In the absence of bias and
magnetic field, two of these bands meet at the six valley points $\mathbf{K},%
\mathbf{K}^{\prime }$ of the reciprocal lattice. Here, we take as the two
non-equivalent points $\mathbf{K}=\left( -4\pi /3a_{0},0\right) $ and $%
\mathbf{K}^{\prime }=-\mathbf{K}=\left( 4\pi /3a_{0},0\right) $ where $%
a_{0}=2.46$ \AA\ is the lattice parameter of graphene. The dispersion of the
two low-energy bands, for small wavevector $\mathbf{k}$ measured from either 
$\mathbf{K}$ or $\mathbf{K}^{\prime },$ is given by $E_{\xi \mathbf{K}%
}\left( \mathbf{k}\right) =\pm \hslash ^{2}k^{2}/2m^{\ast }$ with valley
index $\xi =\pm 1$. The effective electronic mass is defined by $m^{\ast
}=2\hslash ^{2}\gamma _{1}/3\gamma _{0}^{2}a_{0}^{2}=0.054m_{0}$ where $%
m_{0} $ is the bare electronic mass. The two high-energy bands are separated
from the two low-energy bands by a gap of order $\gamma _{1}$.

To describe the low-energy ($E<<\gamma _{1}\,$) excitations of the
tight-binding model, we assume complete spin polarization and use an
effective two-band model\cite{Mccann} where the Hamiltonian is given by%
\begin{equation}
H_{\xi \mathbf{K}}^{0}=\left( 
\begin{array}{cc}
-\xi \frac{\Delta _{B}}{2}+\left( \zeta +\xi \beta \Delta _{B}\right)
aa^{\dag } & \hslash \omega _{c}^{\ast }a^{2} \\ 
\hslash \omega _{c}^{\ast }\left( a^{\dag }\right) ^{2} & \xi \frac{\Delta
_{B}}{2}+\left( \zeta -\xi \beta \Delta _{B}\right) a^{\dag }a%
\end{array}%
\right)  \label{s1}
\end{equation}%
in the basis $\left( A_{2},B_{1}\right) $ for $H_{\mathbf{K}}^{0}$ and $%
\left( B_{1},A_{2}\right) $ for $H_{-\mathbf{K}}^{0}$. In Eq. (\ref{s1}), $%
a,a^{\dag }$ are the ladder operators and we have defined the parameters%
\begin{equation}
\zeta =2\mathrm{sgn}\left( \gamma _{0}\gamma _{4}\right) \sqrt{\beta \beta
_{4}}\gamma _{1}+\beta \Delta ,
\end{equation}%
where sgn denotes the signum function and 
\begin{eqnarray}
\beta &=&\frac{\hslash \omega _{c}^{\ast }}{\gamma _{1}}=8.\,\allowbreak
86\times 10^{-3}B, \\
\beta _{4} &=&\left( \frac{\gamma _{4}}{\gamma _{0}}\right) ^{2}\frac{%
\hslash \omega _{c}^{\ast }}{\gamma _{1}}=1.\,\allowbreak 31\times 10^{-5}B,
\end{eqnarray}%
(with $B$ in Tesla). The effective cyclotron frequency is defined by $\omega
_{c}^{\ast }=eB/m^{\ast }c.$

When $\gamma _{4}=\Delta =\Delta _{B}=0,$ the Landau level energies in each
valley are given by%
\begin{equation}
E_{\xi \mathbf{K}}^{0}=\pm \sqrt{\left\vert N\right\vert \left( \left\vert
N\right\vert +1\right) }\hslash \omega _{c}^{\ast },
\end{equation}%
with $N=0,\pm 1,\pm 2,...$ where $N$ is the Landau level. If the Zeeman
coupling is taken as zero, all Landau levels are thus four time degenerate
(including spin and valley degrees of freedom) with the exception of $N=0$
that is eight times degenerate as shown below. With finite values of $\gamma
_{4}$, $\Delta $ or $\Delta _{B}$, the degeneracy is lifted and we find for
the states in $N=0$ the following spinors and energies (we use the Landau
gauge with $\mathbf{A}=\left( 0,Bx,0\right) $): 
\begin{eqnarray}
\left( 
\begin{array}{c}
0 \\ 
h_{0,X,\sigma }\left( \mathbf{r}\right)%
\end{array}%
\right) ,\;E_{\xi \mathbf{K},0,X,\sigma }^{0} &=&\frac{1}{2}\xi \Delta
_{B}-\sigma \Delta _{z},  \label{s10} \\
\left( 
\begin{array}{c}
0 \\ 
h_{1,X,\sigma }\left( \mathbf{r}\right)%
\end{array}%
\right) ,\;E_{\xi \mathbf{K},1,X,\sigma }^{0} &=&\frac{1}{2}\xi \Delta
_{B}-\sigma \Delta _{z}-\xi \beta \Delta _{B}+\zeta ,  \notag
\end{eqnarray}%
where we have added a Zeeman coupling $\Delta _{z}=g\mu _{B}B/2=0.58\times
10^{-4}B$ eV (with $B$ in Tesla) for more generality. In the absence of
couplings and with $\zeta =0,$ the $N=0$ Landau level has an extra orbital
degeneracy since the two spinors above are then degenerate. The wave
functions $h_{n,X}\left( \mathbf{r}\right) =e^{-iXy/\ell ^{2}}\varphi
_{n}\left( x-X\right) /\sqrt{L_{y}}$ are the eigenstates of an electron with
guiding center $X$ in the Landau gauge, and $\varphi _{n}\left( x\right) $
is the wave function of the one-dimensional harmonic oscillator. Note that
with our choice of normalization for the functions $\varphi _{n}\left(
x\right) $, the action of the ladder operators on the states $\varphi
_{n}\left( x\right) $ is given by $a^{\dag }\varphi _{n}\left( x\right) =i%
\sqrt{n+1}\varphi _{n+1}\left( x\right) $ and $a\varphi _{n}\left( x\right)
=-i\sqrt{n}\varphi _{n-1}\left( x\right) .$

We only consider Landau level $N=0$ in our work so that, from now on, we
will drop the index $N.$ The index $n$ will always refer to the orbital
states $n=0,1$ i.e. to $\varphi _{0}\left( x\right) $ and $\varphi
_{1}\left( x\right) .$ Note that it is clear from Eq. (\ref{s10}) that the
valley $\mathbf{K}(\mathbf{K}^{\prime })$ eigenstates are localized in the
top(bottom) layer. For $N=0$, the layer index is thus equivalent to the
valley index.

\section{QUANTUM\ HALL\ FERROMAGNETS}

To describe the electronic phases in Landau level $N=0$, we use a pseudospin
language where we associate to the layer $\xi =\pm \mathbf{K}$ (or valley)
and orbitals $n=0,1$ a spin-half pseudospin. When Coulomb interaction is
considered in the Hartree-Fock approximation, the ground states at zero bias
and $\zeta =0$ obey a set of Hund's rules in which spin polarization is
maximized first, then layer polarization to the greatest extent possible,
and finally orbital polarization to the extent allowed by the first two rules%
\cite{BarlasPRL1}. In the pseudospin language, the ground states are thus
various types of quantum Hall ferromagnets (QHF's).

At zero bias and with $\zeta =0$, interlayer coherence is present in the
ground state at all integer filling factors but there is no orbital or spin
coherence. Interlayer coherence occurs naturally because of the proximity of
the two layers (the interlayer separation $d=3.34$ \AA\ is very small in
bilayer graphene) but disappears quickly when a finite bias $\Delta _{B}$ is
applied. Adding a finite bias $\Delta _{B}$ enriches considerably the phase
diagram with new states such as orbital and spin QHF's.

In previous works\cite{BarlasPRL1,CoteOrbital}, it was found that above a
critical bias $\Delta _{B}\left( \nu \right) $, the ground state is an
orbital QHF with Ising character (i.e. no orbital coherence) for $\nu =-3$
and a ground state with an $x-y$ character (or orbital coherence) for $\nu
=-1.$ The dispersion relation of the orbital pseudospin mode is gapped at $%
\omega \left( \mathbf{q}=0\right) =\beta \Delta _{B}$ in the former case and
gapless with an highly anisotropic dispersion at small wave vector in the
latter case. To get these results, it was assumed that complete spin
polarization holds even at finite bias and the correction $\zeta $ was not
considered in the single-particle energies. When these two assumptions are
relaxed, the phase diagram at $\Delta _{B}\neq 0$ is modified. An exhaustive
study of the phase diagram for uniform states at all integer filling factors 
$\nu \in \left[ -3,4\right] $ will be presented elsewhere\cite{Lambert}. We
want to mention here, however, some changes that occur in the phase diagram.

\begin{figure}[tbph]
\includegraphics[scale=1]{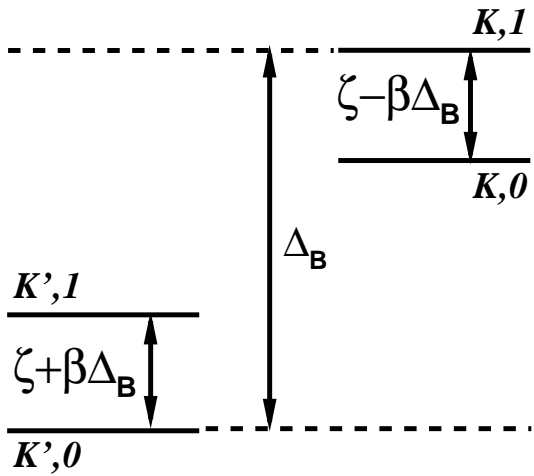}
\caption{Energy levels for one spin state in the presence of a finite bias.}
\label{figniveaux}
\end{figure}

First of all, we remark that orbital coherence is driven by the $\beta
\Delta _{B}$ term in Eq. (\ref{s10}). Together with the term $\zeta ,$ they
lift the degeneracy between levels $n=0$ and $n=1.$The ordering of the
non-iteracting levels is then as illustrated in Fig. \ref{figniveaux}.

When the correction $\zeta $ is neglected, the orbital splitting $\beta
\Delta _{B}<<\Delta _{B}$. For positive bias, the $n=0$ orbital state in the
bottom(top) layer is lower(higher) in energy than the $n=1$ orbital state.
The orbital energy splitting has thus different signs in the $\mathbf{K}\ $%
and $\mathbf{K}^{\prime }$ valleys. No orbital coherence is possible at $\nu
=-3$ because level $n=1$ is always above level $n=0$ at all bias. In that
case, the ground state has all electrons in valley $\mathbf{K}^{\prime }$
with $n=0$ because the Coulomb exchange energy is more negative in $n=0$
than in $n=1.$ The situation is different for $\nu =-1$. In that case, the
valley $\mathbf{K}^{\prime }$ is filled above a critical bias and the
remaining electrons occupy valley $\mathbf{K}$ where level $n=1$ is now
below $n=0$. Now, because of the better exchange interaction in $n=0,$ the
eigenstates in $\mathbf{K}$ are bonding and anti-bonding states of $n=0$ and 
$n=1$ with the electrons mostly in $n=0$ at low bias and mostly in $n=1$ at
large bias. This produces an orbital coherent state.

If $\zeta \neq 0$, we see from Eq. (\ref{s10}) that the only effect is to
push the orbital coherent states to higher bias. In itself, this is not
dramatic provided our effective two-band model is still valid at the new
critical bias. But, if the bias is increased, the spin degree of freedom
must be considered. At zero bias, the four spin up states are separated in
energy from the four spin down states by an exchange-enhanced Zeeman gap of
order $e^{2}/\kappa \ell $ where $\kappa $ is the dielectric constant of the
substrate and $\ell =\sqrt{\hslash c/eB}$ is the magnetic length. With bias,
two levels with spin up(down) are shifted upward(downward) in energy. When
levels with different spins cross, states with spin coherence ($x-y$ spin
QHF) become possible and they may replace the orbital state in the phase
diagram. Whether this is the case or not must be decided by numerical
calculations. For $\gamma _{4}=0.12,$ a numerical calculation\cite{Lambert}
shows that the orbital coherent state is absent for $\nu =-1$. We remark,
however, that the exact value of $\gamma _{4}$ is not precisely known and
than orbital coherence does occur at smaller values of this parameter.

Fortunately, the phase diagram for the eight-level system is very rich and
numerical calculations\cite{Lambert} show that orbital coherence with no
interlayer or spin coherences occurs in some range of bias but at the higher
filling factors $\nu =1$ and $\nu =3$ and even when $\gamma _{4}$ is as
large as $\gamma _{4}=0.12$ and spin mixing is considered. Since we are
confident that orbital coherent states do occur in the phase diagram in
bilayer graphene, we will study, in this paper, the simplest case with $\nu
=-1$,$\gamma _{4}=0$ and complete spin polarization. Our results should
apply with some minor changes to the orbital states at $\nu =1$ and $\nu =3.$

In concluding this section, we remark that the phase diagram for filling
factors $\nu =-1$ has been studied in some details before \cite{BarlasPRL1},%
\cite{CoteOrbital} but only the homogeneous states have been considered.
Non-uniform states such as Skyrmion crystals with valley or orbital
pseudospin textures have also been considered but near integer filling
factors only\cite{CoteCrystal}. In this work, our focus is on the \textit{%
non-uniform} states at precisely $\nu =-1.$

\section{FORMALISM\qquad}

In the rest of this paper, we concentrate on the study of the orbital
coherent states at $\nu =-1$, assuming $\zeta =0$ and complete spin
polarization. Our calculations are all done at zero temperature.

\subsection{Hartree-Fock Hamiltonian}

With the restrictions outlined above, the analysis is reduced to that of a
two-level system in valley $\mathbf{K}$ since the two filled levels in
valley $\mathbf{K}^{\prime }$ can be considered as inert. The
non-interacting Hamiltonian is given by Eq. (\ref{s1}) while the many-body
Hamiltonian in the Hartree-Fock approximation is\cite{CoteOrbital}%
\begin{gather}
H_{HF}=N_{\varphi }E_{n}\rho _{n,n}\left( 0\right)  \label{HHF} \\
+N_{\varphi }\overline{\sum_{\mathbf{q}}}H_{n_{1},n_{2},n_{3},n_{4}}\left( 
\mathbf{q}\right) \left\langle \rho _{n_{1},n_{2}}\left( -\mathbf{q}\right)
\right\rangle \rho _{n_{3},n_{4}}\left( \mathbf{q}\right)  \notag \\
-N_{\varphi }\sum_{\mathbf{q}}X_{n_{1},n_{4},n_{3},n_{2}}\left( \mathbf{q}%
\right) \left\langle \rho _{n_{1},n_{2}}\left( -\mathbf{q}\right)
\right\rangle \rho _{n_{3},n_{4}}\left( \mathbf{q}\right) ,  \notag
\end{gather}%
where $n_{i}=0,1$ is the orbital index and $N_{\varphi }=S/2\pi \ell ^{2}$
is the Landau level degeneracy. The single-particle energies can be
simplified to $E_{n}=-n\beta \Delta _{B}$. In deriving Eq. (\ref{HHF}), we
have taken into account a neutralizing positive background so that the $%
\mathbf{q}=0$ contribution is absent in the Hartree term. This convention is
indicated by the bar over the summation.

The density operators in Eq. (\ref{HHF}), are defined by

\begin{eqnarray}
\rho _{n_{1},n_{2}}\left( \mathbf{q}\right) &=&\frac{1}{N_{\varphi }}%
\sum_{X_{1},X_{2}}e^{-\frac{i}{2}q_{x}\left( X_{1}+X_{2}\right) } \\
&&\times c_{X_{1},n_{1}}^{\dagger }c_{X_{2},n_{2}}\delta
_{X_{1},X_{2}+q_{y}\ell ^{2}},  \notag
\end{eqnarray}%
where $c_{X,n}^{\dagger }\left( c_{X,n}\right) $ creates(destroys) an
electron in state $\left( X,n\right) $ in the Landau gauge. The Hartree and
Fock interactions are given by 
\begin{eqnarray}
H_{n_{1},n_{2},n_{3},n_{4}}\left( \mathbf{q}\right) &=&\left( \frac{e^{2}}{%
\kappa \ell }\right) \frac{1}{q\ell }K_{n_{1},n_{2}}\left( \mathbf{q}\right)
K_{n_{3},n_{4}}\left( -\mathbf{q}\right) ,  \label{hartree1} \\
X_{n_{1},n_{2},n_{3},n_{4}}\left( \mathbf{q}\right) &=&\int \frac{d\mathbf{p}%
\ell ^{2}}{2\pi }H_{n_{1},n_{2},n_{3},n_{4}}\left( \mathbf{p}\right) e^{i%
\mathbf{q}\times \mathbf{p}\ell ^{2}},  \label{hartree2}
\end{eqnarray}%
where $\kappa $ is the effective dielectric constant at the position of the
graphene layers. The Coulomb energy $e^{2}/\kappa \ell =1\,\allowbreak 1.3%
\sqrt{B}$ meV with $B$ in Tesla and we take $\kappa =5.$

The form factors which appear in $H$ and $X$ are given by%
\begin{eqnarray}
K_{0,0}\left( \mathbf{q}\right) &=&\exp \left( \frac{-q^{2}\ell ^{2}}{4}%
\right) ,  \label{form1} \\
K_{1,1}\left( \mathbf{q}\right) &=&\exp \left( \frac{-q^{2}\ell ^{2}}{4}%
\right) \left( 1-\frac{q^{2}\ell ^{2}}{2}\right) ,  \notag  \label{form2} \\
K_{1,0}\left( \mathbf{q}\right) &=&\left( \frac{\left( q_{y}+iq_{x}\right)
\ell }{\sqrt{2}}\right) \exp \left( \frac{-q^{2}\ell ^{2}}{4}\right) , 
\notag  \label{form3} \\
K_{0,1}\left( \mathbf{q}\right) &=&\left( \frac{\left( -q_{y}+iq_{x}\right)
\ell }{\sqrt{2}}\right) \exp \left( \frac{-q^{2}\ell ^{2}}{4}\right) . 
\notag  \label{form4}
\end{eqnarray}%
They capture the character of the two different orbital states and play an
important role in the physics of the orbital phase. Detailed expressions for
the Hartree and Fock interactions $H$ and $X$ are given in Appendix A of
Ref. \onlinecite{CoteOrbital}.

\subsection{Order parameters in the coherent phases}

The order parameters of the orbital phases are obtained from the
single-particle Green's function 
\begin{equation}
G_{n,n^{\prime }}\left( X,X^{\prime },\tau \right) =-\left\langle T_{\tau
}c_{n,X}\left( \tau \right) c_{n^{\prime },X^{\prime }}^{\dagger }\left(
0\right) \right\rangle ,  \label{GFX}
\end{equation}%
where $T_{\tau }$ is the imaginary time ordering operator. We define the
Fourier transform of the single-particle Green's function as

\begin{eqnarray}
G_{n,n^{\prime }}\left( \mathbf{q,}\tau \right) &=&\frac{1}{N_{\varphi }}%
\sum_{X,X^{\prime }}e^{-\frac{i}{2}q_{x}\left( X+X^{\prime }\right) }
\label{GF} \\
&&\times \delta _{X,X^{\prime }-q_{y}\ell ^{2}}G_{n,n^{\prime }}\left(
X,X^{\prime },\tau \right)  \notag
\end{eqnarray}%
so that the order parameters of the coherent phases are related to the
Green's function by 
\begin{equation}
\left\langle \rho _{n,n^{\prime }}\left( \mathbf{q}\right) \right\rangle
=G_{n^{\prime },n}\left( \mathbf{q,}\tau =0^{-}\right) .
\end{equation}%
The equation of motion for the Green's function in the Matsubara formalism
and in the Hartree-Fock approximation is given by\newline
\begin{gather}
\left( i\omega _{n}+\mu /\hslash \right) G_{n,n^{\prime }}\left( \mathbf{q}%
,i\omega _{n}\right) -  \label{EHF} \\
\sum_{m=0,1}\sum_{\mathbf{q}^{\prime }}T_{n,m}\left( \mathbf{q},\mathbf{q}%
^{\prime }\right) \gamma _{\mathbf{q},\mathbf{q}^{\prime }}G_{m,n^{\prime
}}\left( \mathbf{q},i\omega _{n}\right) =\delta _{n,n^{\prime }}\delta _{%
\mathbf{q},0},  \notag
\end{gather}%
with the matrix elements%
\begin{eqnarray}
T_{n,m}\left( \mathbf{q},\mathbf{q}^{\prime }\right) &=&U_{n,m}^{H}\left( 
\mathbf{q-q}^{\prime }\right) -U_{n,m}^{F}\left( \mathbf{q-q}^{\prime
}\right) \\
&&-\beta \Delta _{B}\delta _{\mathbf{q},\mathbf{q}^{\prime }}\delta
_{n,m}\delta _{n,1},  \notag
\end{eqnarray}%
and the phase factor 
\begin{equation}
\gamma _{\mathbf{q},\mathbf{q}^{\prime }}=e^{-i\mathbf{q}\times \mathbf{q}%
^{\prime }\ell ^{2}/2}.
\end{equation}%
The Hartree and Fock potentials are defined by%
\begin{eqnarray}
U_{n_{3},n_{4}}^{H}\left( \mathbf{q}\right)
&=&\sum_{n_{1},n_{2}}H_{n_{1},n_{2},n_{3},n_{4}}\left( -\mathbf{q}\right)
\left\langle \rho _{n_{1},n_{2}}\left( \mathbf{q}\right) \right\rangle , \\
U_{n_{3},n_{4}}^{F}\left( \mathbf{q}\right)
&=&\sum_{n_{1},n_{2}}X_{n_{1},n_{4},n_{3},n_{2}}\left( -\mathbf{q}\right)
\left\langle \rho _{n_{1},n_{2}}\left( \mathbf{q}\right) \right\rangle .
\end{eqnarray}

The self-consistent Eq. (\ref{EHF}) must be solved numerically in an
iterative way in order to get the order parameters in the different orbital
phases. Once this is done, the Hartree-Fock energy is obtained from 
\begin{align}
& \frac{E_{HF}}{N}=-\beta \Delta _{B}\left\langle \rho _{1,1}\left( 0\right)
\right\rangle  \label{EHFN} \\
& +\frac{1}{2}\overline{\sum_{\mathbf{q}}}%
\sum_{n_{1},...,n_{4}}H_{n_{1},n_{2},n_{3},n_{4}}\left( \mathbf{q}\right)
\left\langle \rho _{n_{1},n_{2}}\left( -\mathbf{q}\right) \right\rangle
\left\langle \rho _{n_{3},n_{4}}\left( \mathbf{q}\right) \right\rangle 
\notag \\
& -\frac{1}{2}\sum_{\mathbf{q}%
}\sum_{n_{1},...,n_{4}}X_{n_{1},n_{4},n_{3},n_{2}}\left( \mathbf{q}\right)
\left\langle \rho _{n_{1},n_{2}}\left( -\mathbf{q}\right) \right\rangle
\left\langle \rho _{n_{3},n_{4}}\left( \mathbf{q}\right) \right\rangle . 
\notag
\end{align}

The Hartree-Fock equation of motion for the Green's function leads to the
sum rule (at $T=0$K):%
\begin{equation}
\sum_{\mathbf{q}}\sum_{n_{2}}\left\vert \left\langle \rho
_{n_{1},n_{2}}\left( \mathbf{q}\right) \right\rangle \right\vert
^{2}=\left\langle \rho _{n_{1},n_{1}}\left( 0\right) \right\rangle ,
\label{SR}
\end{equation}%
and we have by definition%
\begin{equation}
\left\langle \rho _{n,n}\left( 0\right) \right\rangle =\nu _{n},
\end{equation}%
where $\nu _{n}$ is the filling factor of the $n$ level.

\subsection{Induced electric dipoles}

In the pseudospin language, we associate state $n=0$ with pseudospin up and $%
n=1$ with pseudospin down. We then define the \textit{guiding-center }%
density operator%
\begin{equation}
\rho _{n}\left( \mathbf{q}\right) =\rho _{0,0}\left( \mathbf{q}\right) +\rho
_{1,1}\left( \mathbf{q}\right) ,
\end{equation}%
and the three components of the pseudospin density vector $\overrightarrow{%
\rho }\left( \mathbf{q}\right) $ are given by 
\begin{eqnarray}
\rho _{x}\left( \mathbf{q}\right) &=&\frac{1}{2}\left( \rho _{0,1}\left( 
\mathbf{q}\right) +\rho _{1,0}\left( \mathbf{q}\right) \right) ,
\label{pseudospin} \\
\rho _{y}\left( \mathbf{q}\right) &=&\frac{1}{2i}\left( \rho _{0,1}\left( 
\mathbf{q}\right) -\rho _{1,0}\left( \mathbf{q}\right) \right) ,  \notag \\
\rho _{z}\left( \mathbf{q}\right) &=&\frac{1}{2}\left( \rho _{0,0}\left( 
\mathbf{q}\right) -\rho _{1,1}\left( \mathbf{q}\right) \right) .  \notag
\end{eqnarray}

The coupling of the 2DEG with a uniform \textit{external} electric field in
the plane of the layers is given by $H_{E}=-e\int d\mathbf{r}n\left( \mathbf{%
r}\right) \phi \left( \mathbf{r}\right) $ with $\mathbf{E}=-\nabla \phi
\left( \mathbf{r}\right) .$ In this expression, we must use the true density 
$n\left( \mathbf{r}\right) $ which is the Fourier transform of%
\begin{equation}
n\left( \mathbf{q}\right) =N_{\varphi }\sum_{n_{1},n_{2}}\rho
_{n_{1},n_{2}}\left( \mathbf{q}\right) K_{n_{1},n_{2}}\left( -\mathbf{q}%
\right) .  \label{realdensity}
\end{equation}%
This coupling can be written as%
\begin{eqnarray}
H &=&-eN_{\varphi }\int d\mathbf{r}\overline{\rho }_{n}\left( \mathbf{r}%
\right) \phi \left( \mathbf{r}\right) \\
&&+\sqrt{2}\ell eN_{\varphi }\left( \rho _{x}\left( \mathbf{q}=0\right)
E_{x}-\rho _{y}\left( \mathbf{q}=0\right) E_{y}\right) ,  \notag
\end{eqnarray}%
where we have defined $\overline{\rho }_{i}\left( \mathbf{q}\right) =\exp
\left( -q^{2}\ell ^{2}/4\right) \rho _{i}\left( \mathbf{q}\right) $ with $%
i=n,x,y,z.$ In the three states studied in this paper (uniform, helical, and
Skyrme crystal), the guiding-center density $\left\langle \rho _{n}\left( 
\mathbf{r}\right) \right\rangle $ is uniform and so $\left\langle \rho
_{n}\left( \mathbf{q}\right) \right\rangle =\delta _{\mathbf{q},0}.$ For
these states, the coupling with the electric field is simply given by%
\begin{equation}
H_{E}=-N_{\varphi }\mathbf{d}\left( 0\right) \cdot \mathbf{E,}
\end{equation}%
where we have defined the electric dipole operator%
\begin{equation}
\mathbf{d}\left( \mathbf{q}\right) =-e\sqrt{2}\ell e^{\frac{-q^{2}\ell ^{2}}{%
4}}\left( \overline{\rho }_{x}\left( \mathbf{q}\right) ,-\overline{\rho }%
_{y}\left( \mathbf{q}\right) \right) .
\end{equation}%
The fact that orbital coherence leads to a finite density of electric
dipoles in the plane of the layers was first shown in Ref. %
\onlinecite{Shizuya1}.

Note that for the states with $\left\langle \rho _{n}\left( \mathbf{q}%
\right) \right\rangle =\delta _{\mathbf{q},0}$, the sum rules of Eq. (\ref%
{SR}) are, in pseudospin language, equivalent to

\begin{equation}
\sum_{\mathbf{q}}\left\vert \left\langle \rho _{x}\left( \mathbf{q}\right)
\right\rangle \right\vert ^{2}+\left\vert \left\langle \rho _{y}\left( 
\mathbf{q}\right) \right\rangle \right\vert ^{2}+\left\vert \left\langle
\rho _{z}\left( \mathbf{q}\right) \right\rangle \right\vert ^{2}=\frac{1}{4}
\label{SR2}
\end{equation}%
which is an \textit{average} normalization condition for the pseudospin
field vector i.e. $\int d\mathbf{r}\left\langle \mathbf{\rho }\left( \mathbf{%
r}\right) \right\rangle ^{2}=1/4S$ where $S$ is the area of the 2DEG. The
modulus of the pseudospin field is, in general, not constant in space.

\subsection{Collective modes}

To study the collective excitations, we compute the two-particle Green's
function

\begin{align}
& \chi _{n_{1},n_{2},n_{3},n_{4}}\left( \mathbf{q},\mathbf{q}^{\prime };\tau
\right)  \label{twopart} \\
& =-N_{\varphi }\left\langle T_{\tau }\rho _{n_{1},n_{2}}\left( \mathbf{q,}%
\tau \right) \rho _{n_{3},n_{4}}\left( -\mathbf{q}^{\prime },0\right)
\right\rangle  \notag \\
& +N_{\varphi }\left\langle \rho _{n_{1},n_{2}}\left( \mathbf{q}\right)
\right\rangle \left\langle \rho _{n_{3},n_{4}}\left( -\mathbf{q}^{\prime
}\right) \right\rangle  \notag
\end{align}%
in the generalized random-phase approximation (GRPA). In this approximation, 
$\chi _{n_{1},n_{2},n_{3},n_{4}}\left( \mathbf{q},\mathbf{q}^{\prime };\tau
\right) $ is the solution of the equation

\begin{eqnarray}
&&\chi _{n_{1},n_{2},n_{3},n_{4}}\left( \mathbf{q},\mathbf{q}^{\prime
};i\Omega _{n}\right)  \label{grpa1} \\
&=&\chi _{n_{1},n_{2},n_{3},n_{4}}^{0}\left( \mathbf{q},\mathbf{q}^{\prime
};i\Omega _{n}\right)  \notag \\
&&+\frac{1}{\hslash }\sum_{n_{5},...,n_{8}}\sum_{\mathbf{q}^{\prime \prime
}}\chi _{n_{1},n_{2},n_{5},n_{6}}^{0}\left( \mathbf{q},\mathbf{q}^{\prime
\prime };i\Omega _{n}\right)  \notag \\
&&\times H_{n_{5},n_{6},n_{7},n_{8}}\left( \mathbf{q}^{\prime \prime
}\right) \chi _{n_{7},n_{8},n_{3},n_{4}}\left( \mathbf{q}^{\prime \prime },%
\mathbf{q}^{\prime };i\Omega _{n}\right)  \notag \\
&&-\frac{1}{\hslash }\sum_{n_{5},...,n_{8}}\sum_{\mathbf{q}^{\prime \prime
}}\chi _{n_{1},n_{2},n_{5},n_{6}}^{0}\left( \mathbf{q},\mathbf{q}^{\prime
\prime };i\Omega _{n}\right)  \notag \\
&&\times X_{n_{5},n_{8},n_{7},n_{6}}\left( \mathbf{q}^{\prime \prime
}\right) \chi _{n_{7},n_{8},n_{3},n_{4}}\left( \mathbf{q}^{\prime \prime },%
\mathbf{q}^{\prime };i\Omega _{n}\right) ,  \notag
\end{eqnarray}%
where $\Omega _{n}$ is a bosonic Matsubura frequency and the Hartree-Fock
two-particle Green's function $\chi _{n_{1},n_{2},n_{3},n_{4}}^{0}\left( 
\mathbf{q},\mathbf{q}^{\prime };i\Omega _{n}\right) $ is given by

\begin{eqnarray}
&&\left[ i\hslash \Omega _{n}-\left( E_{n_{2}}-E_{n_{1}}\right) \right] \chi
_{n_{1},n_{2},n_{3},n_{4}}^{0}\left( \mathbf{q},\mathbf{q}^{\prime },\Omega
_{n}\right)  \label{grpa2} \\
&=&\hslash \left\langle \rho _{n_{1},n_{4}}\left( \mathbf{q-q}^{\prime
}\right) \right\rangle \delta _{n_{2},n_{3}}\gamma _{\mathbf{q},\mathbf{q}%
^{\prime }}^{\ast }  \notag \\
&&-\hslash \left\langle \rho _{n_{3},n_{2}}\left( \mathbf{q-q}^{\prime
}\right) \right\rangle \delta _{n_{1},n_{4}}\gamma _{\mathbf{q},\mathbf{q}%
^{\prime }}  \notag \\
&&-\sum_{n}\overline{\sum_{\mathbf{q}^{\prime \prime }}}U_{n,n_{1}}^{H}%
\left( \mathbf{q-q}^{\prime \prime }\right) \gamma _{\mathbf{q},\mathbf{q}%
^{\prime \prime }}^{\ast }\chi _{n,n_{2},n_{3},n_{4}}^{0}\left( \mathbf{q}%
^{\prime \prime },\mathbf{q}^{\prime },\Omega _{n}\right)  \notag \\
&&+\sum_{n}\overline{\sum_{\mathbf{q}^{\prime \prime }}}U_{n_{2},n}^{H}%
\left( \mathbf{q-q}^{\prime \prime }\right) \gamma _{\mathbf{q},\mathbf{q}%
^{\prime \prime }}\chi _{n_{1},n,n_{3},n_{4}}^{0}\left( \mathbf{q}^{\prime
\prime },\mathbf{q}^{\prime },\Omega _{n}\right)  \notag \\
&&+\sum_{n}\sum_{\mathbf{q}^{\prime \prime }}U_{n,n_{1}}^{F}\left( \mathbf{%
q-q}^{\prime \prime }\right) \gamma _{\mathbf{q},\mathbf{q}^{\prime \prime
}}^{\ast }\chi _{n,n_{2},n_{3},n_{4}}^{0}\left( \mathbf{q}^{\prime \prime },%
\mathbf{q}^{\prime },\Omega _{n}\right)  \notag \\
&&-\sum_{n}\sum_{\mathbf{q}^{\prime \prime }}U_{n_{2},n}^{F}\left( \mathbf{%
q-q}^{\prime \prime }\right) \gamma _{\mathbf{q},\mathbf{q}^{\prime \prime
}}\chi _{n_{1},n,n_{3},n_{4}}^{0}\left( \mathbf{q}^{\prime \prime },\mathbf{q%
}^{\prime },\Omega _{n}\right) .  \notag
\end{eqnarray}%
Note that the response functions depend only on the order parameters $%
\left\langle \rho _{n,m}\left( \mathbf{q}\right) \right\rangle $ computed in
the HFA. Eqs. (\ref{grpa1},\ref{grpa2}) can be solved numerically by writing
them in a matrix form. The collective excitations are then given by the
poles of the retarded Green's function $\chi
_{n_{1},n_{2},n_{3},n_{4}}^{R}\left( \mathbf{q},\mathbf{q}^{\prime },\omega
\right) $ which is obtained by the analytic continuation $i\Omega
_{n}\rightarrow \omega +i\delta $ of the corresponding two-particle Green's
function.

\section{PHASE\ DIAGRAM\ FOR THE ORBITAL PHASES AT $\protect\nu =-1$}

\begin{figure}[tbph]
\includegraphics[scale=1]{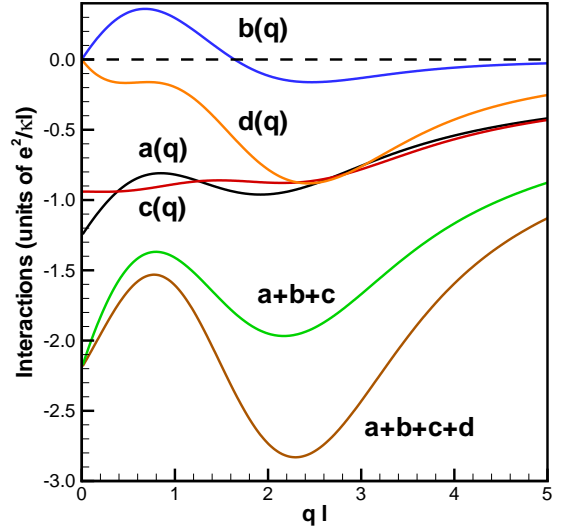}
\caption{(Color online) Effective pseudospin interactions as a function of
wave vector $q\ell .$}
\label{interactions}
\end{figure}

We define an average pseudospin field by%
\begin{equation}
\mathbf{p}\left( \mathbf{q}\right) =\sqrt{2}\ell e\left( -\left\langle \rho
_{x}\left( \mathbf{q}\right) \right\rangle ,\left\langle \rho _{y}\left( 
\mathbf{q}\right) \right\rangle ,\left\langle \rho _{z}\left( \mathbf{q}%
\right) \right\rangle \right) ,
\end{equation}%
so that the dipole field can be related to the parallel component of this
vector by%
\begin{equation}
\mathbf{d}\left( \mathbf{q}\right) =e^{\frac{-q^{2}\ell ^{2}}{4}}\mathbf{p}%
_{\Vert }\left( \mathbf{q}\right) .  \label{dipole}
\end{equation}%
After a lengthy calculation, we can put the Hartree-Fock energy per electron
given by Eq. (\ref{EHFN}) in the pseudospin form 
\begin{eqnarray}
&&\frac{E_{HF}}{N}/\left( \frac{e^{2}}{\kappa \ell }\right)  \label{EPP} \\
&=&-\frac{11}{32}\sqrt{\frac{\pi }{2}}-\frac{1}{2}\beta \overline{\Delta }%
_{B}  \notag \\
&&+\frac{1}{\alpha }\beta \left( \overline{\Delta }_{B}-\frac{1}{2}\overline{%
\Delta }_{B}^{(1)}\right) p_{z}\left( 0\right)  \notag \\
&&+\frac{1}{2\alpha ^{2}}\sum_{\mathbf{q}}\mathbf{p}_{\Vert }\left( \mathbf{%
-q}\right) \cdot \left[ a\left( q\right) \mathbf{I}+b\left( q\right) \mathbf{%
\Lambda }\left( \mathbf{q}\right) \right] \cdot \mathbf{p}_{\Vert }\left( 
\mathbf{q}\right)  \notag \\
&&+\frac{1}{2\alpha ^{2}}\sum_{\mathbf{q}}c\left( q\right) p_{z}\left( -%
\mathbf{q}\right) p_{z}\left( \mathbf{q}\right)  \notag \\
&&+\frac{i}{4\alpha ^{2}}\sum_{\mathbf{q}}d\left( q\right) \left( \widehat{%
\mathbf{z}}\times \widehat{\mathbf{q}}\right) \cdot \left( \mathbf{p}\left( 
\mathbf{-q}\right) \times \mathbf{p}\left( \mathbf{q}\right) \right) . 
\notag
\end{eqnarray}%
In this equation, the bias $\overline{\Delta }_{B}=\Delta _{B}/\left(
e^{2}/\kappa \left( \ell \right) \right) ,\alpha =\sqrt{2}\ell e,$ $\mathbf{I%
}$ is the $2\times 2$ unit matrix and $\mathbf{\Lambda }\left( \mathbf{q}%
\right) $ is given by 
\begin{equation}
\mathbf{\Lambda }\left( \mathbf{q}\right) =\left( 
\begin{array}{cc}
\cos \left( 2\varphi _{\mathbf{q}}\right) & \sin \left( 2\varphi _{\mathbf{q}%
}\right) \\ 
\sin \left( 2\varphi _{\mathbf{q}}\right) & -\cos \left( 2\varphi _{\mathbf{q%
}}\right)%
\end{array}%
\right) ,
\end{equation}%
where $\varphi _{\mathbf{q}}$ is the angle between the wave vector $\mathbf{q%
}$ and the $x$ axis. The bias $\overline{\Delta }_{B}^{(1)}/2$ is defined as
the bias for which the charge is equally distributed between the two levels $%
n=0,1$ in the uniform phase. Finally, the pseudospin interactions are given
by%
\begin{eqnarray}
a\left( q\right) &=&2\left( H_{0,1,1,0}\left( q\right) -X_{1,1,0,0}\left(
q\right) \right) \\
&=&q\ell e^{-\frac{q^{2}\ell ^{2}}{2}}  \notag \\
&&-2\int_{0}^{\infty }dy\left( 1-\frac{y^{2}}{2}\right)
e^{-y^{2}/2}J_{0}\left( q\ell y\right) ,  \notag
\end{eqnarray}%
\begin{eqnarray}
b\left( q\right) &=&2e^{2i\varphi _{\mathbf{q}}}\left( H_{1,0,1,0}\left( 
\mathbf{q}\right) -X_{1,0,1,0}\left( \mathbf{q}\right) \right)
\label{binter} \\
&=&q\ell e^{-\frac{q^{2}\ell ^{2}}{2}}-\int_{0}^{\infty
}dyy^{2}e^{-y^{2}/2}J_{2}\left( q\ell y\right) ,  \notag
\end{eqnarray}%
\begin{eqnarray}
c\left( q\right) &=&H_{0,0,0,0}\left( q\right) -X_{0,0,0,0}\left( q\right) \\
&&+H_{1,1,1,1}\left( q\right) -X_{1,1,1,1}\left( q\right)  \notag \\
&&-2\left( H_{1,1,0,0}\left( q\right) -X_{0,1,1,0}\left( q\right) \right) 
\notag \\
&=&\frac{q^{3}\ell ^{3}}{4}e^{-\frac{q^{2}\ell ^{2}}{2}}  \notag \\
&&-\int_{0}^{\infty }dy\left( 2-2y^{2}+\frac{y^{4}}{4}\right)
e^{-y^{2}/2}J_{0}\left( q\ell y\right) ,  \notag
\end{eqnarray}

\begin{eqnarray}
d\left( q\right) &=&-4ie^{i\varphi _{\mathbf{q}}}\left( H_{1,0,0,0}\left( 
\mathbf{q}\right) -X_{1,0,0,0}\left( \mathbf{q}\right) \right)  \notag \\
&&-4ie^{i\varphi _{\mathbf{q}}}\left( H_{1,1,1,0}\left( \mathbf{q}\right)
-X_{1,1,1,0}\left( -\mathbf{q}\right) \right) \\
&=&\frac{4}{\sqrt{2}}\frac{q^{2}\ell ^{2}}{2}e^{-\frac{q^{2}\ell ^{2}}{2}} 
\notag \\
&&-\frac{4}{\sqrt{2}}\int_{0}^{\infty }dyy\left( 2-\frac{y^{2}}{2}\right)
e^{-y^{2}/2}J_{1}\left( q\ell y\right) .  \notag
\end{eqnarray}%
These interactions are plotted in Fig. \ref{interactions}. Their values at $%
\mathbf{q}=0$ are $a\left( 0\right) =-\sqrt{\frac{\pi }{2}},b\left( 0\right)
=0,c\left( 0\right) =-\frac{3}{4}\sqrt{\frac{\pi }{2}},d\left( 0\right) =0.$

Eq. (\ref{EPP}) is the energy functional of an effective helical pseudospin
ferromagnet. Apart from the constant terms, there are four distinct
contributions to the total energy:

\begin{enumerate}
\item The term $\beta \left( \overline{\Delta }_{B}-\overline{\Delta }%
_{B}^{(1)}/2\right) p_{z}\left( 0\right) $ is an effective Zeeman coupling
that changes sign at the bias $\overline{\Delta }_{B}^{(1)}/2.$ For $%
\overline{\Delta }_{B}<\overline{\Delta }_{B}^{(1)}/2,$ the pseudospins
order along the positive $z$ axis. They order along $-z$ for $\overline{%
\Delta }_{B}>\overline{\Delta }_{B}^{(1)}/2.$

\item The terms involving $a\left( q\right) $ and $c\left( q\right) $ are
non local Heisenberg exchange interactions between the pseudospins.

\item The term with $b\left( q\right) \mathbf{\Lambda }\left( \mathbf{q}%
\right) $ is a dipolar interaction. In the small wave vector limit, only the
Hartree term $q\ell e^{-\frac{q^{2}\ell ^{2}}{2}}$ in Eq. (\ref{binter})
contributes significatively to $b\left( q\right) $. To first order in $q\ell
,b\left( q\right) \rightarrow q\ell $. The dipolar term can be related to
the Fourier transform of the dipole-dipole electrostatic interaction.

\item The fourth term is a Dzyaloshinskii-Moriya (DM)\ interaction between
the pseudospins. With the redefinition $p_{y}\rightarrow -p_{x}$ and $%
p_{x}\rightarrow p_{y},$ it takes the more conventional DM form which is the
Fourier transform of $D\int d\mathbf{r}\left( \mathbf{p}\cdot \left( \nabla
\times \mathbf{p}\right) \right) $ where $D$ is a constant. In our case, the
DM$\ $constant $D$ becomes a non local function $d\left( \mathbf{r}-\mathbf{r%
}^{\prime }\right) .$ It is interesting to remark that a DM\ occurs in our
model although we are not dealing with real spins or spin-orbit interaction.
The physical origin of this term was discussed in Ref. %
\onlinecite{CoteOrbital}. From a microscopic point of view, $d\left(
q\right) $ contains only the interactions $H_{n_{1},n_{2},n_{3},n_{4}}\left( 
\mathbf{q}\right) $ and $X_{n_{1},n_{2},n_{3},n_{4}}\left( \mathbf{q}\right) 
$ that do not conserve the orbital index (for example: $X_{1,0,0,0}\left( 
\mathbf{q}\right) $)$.$ These terms arises in our system because we are
dealing with two different orbitals $h_{0}\left( \mathbf{r}\right) $ and $%
h_{1}\left( \mathbf{r}\right) .$ On the contrary, the other effective
interactions $a\left( q\right) ,b\left( q\right) $ and $c\left( q\right) $
consist of terms that conserve the orbital index.
\end{enumerate}

The exchange interactions tends to align the pseudospins together while the
DM term favors a rotation of the pseudospins in real space. This type of
competition is usually present in helical magnets such as MnSi and Fe$_{1-x}$%
Co$_{x}$Si. In fact, our numerical calculation gives a phase diagram which
is similar to that found recently in the helical magnet Fe$_{0.5}$Co$_{0.5}$%
Si where Skyrmion crystal and an helical phase have been observed using
Lorentz transmission electron microscopy\cite{XZYu}.

Because the instability of the uniform phase occurs at $q_{y}\ell \approx -2$
(see below), it is not useful to derive a long-wavelength approximation of
Eq. (\ref{EPP}). The calculation of the optimal wave vector $q\ell $ for the
helical state must be done numerically. To obtain the phase diagram for the
orbital phases, we evaluate the Hartree-Fock energy for (once again, $%
\left\langle \rho _{n}\left( \mathbf{q}\right) \right\rangle =\delta _{%
\mathbf{q},0}$ in all three cases):

\begin{enumerate}
\item The uniform phase (UP) defined by $\mathbf{p}\left( \mathbf{q}\right) =%
\mathbf{p}_{0}\delta _{\mathbf{q},0}$

\item The Skyrmion crystal phase (SKP) defined by the set of order
parameters $\left\{ \mathbf{p}\left( \mathbf{G}\right) \right\} $ where $%
\mathbf{G}$ is a reciprocal lattice vector of the crystal. The triangular
crystal has lower energy than the square or rectangular lattices. (We have
not tried other lattice types however.)

\item The helical phase (HP) defined the order parameters $\left\{ \mathbf{p}%
\left( \mathbf{q}\right) \right\} $ where $\mathbf{q}=n\mathbf{q}_{0}$ where 
$n=0,\pm 1,\pm 2,...$
\end{enumerate}

Within this set of states, we find the following ordering:

\begin{equation*}
\begin{tabular}{|l|l|}
\hline
$\overline{\Delta _{B}}<0.06$ & UP \\ \hline
$0.06<\overline{\Delta _{B}}<0.52$ & SKP \\ \hline
$0.52<\overline{\Delta _{B}}<3.02$ & HP \\ \hline
$2.98<\overline{\Delta _{B}}<3.44$ & SKP \\ \hline
$3.44<\overline{\Delta _{B}}<3.536$ & UP \\ \hline
\end{tabular}%
\end{equation*}%
Above $\overline{\Delta _{B}}=\overline{\Delta }_{B}^{(1)}=3.536,$ all
electrons are in level $n=1$ and there is no orbital coherence. We remark
that the phase diagram is symmetrical with respect to the bias $\overline{%
\Delta }_{B}^{(1)}/2$ (see Eq. (\ref{existence}) below). When the layer
index is added to the picture, the ground state is an interlayer coherent
phase with no orbital coherence for $\overline{\Delta _{B}}<0.0022.$ Also,
although the guiding-center density $\left\langle \rho \left( \mathbf{r}%
\right) \right\rangle $ is uniform in space, the real density $n\left( 
\mathbf{r}\right) $ is not. For $\left\langle \rho \left( \mathbf{r}\right)
\right\rangle $ to be constant in space, we must have $\left\langle \rho
_{0,0}\left( \mathbf{q}\right) \right\rangle =-\left\langle \rho
_{1,1}\left( \mathbf{q}\right) \right\rangle $ for $\mathbf{q}\neq 0\mathbf{%
\ }$in the Skyrmion and helical phases.

\subsection{Uniform phase}

The uniform phase has all orbital pseudospins pointing in the same direction
in space. Its energy is given by%
\begin{eqnarray}
&&\frac{E_{HF}}{N}/\left( \frac{e^{2}}{\kappa \ell }\right)  \label{liquidh}
\\
&=&-\frac{11}{32}\sqrt{\frac{\pi }{2}}-\frac{1}{2}\beta \overline{\Delta }%
_{B}  \notag \\
&&+\frac{1}{\alpha }\beta \left( \overline{\Delta }_{B}-\frac{1}{2}\overline{%
\Delta }_{B}^{(1)}\right) p_{z}\left( 0\right)  \notag \\
&&-\frac{1}{2\alpha ^{2}}\sqrt{\frac{\pi }{2}}\left( \mathbf{p}_{\Vert
}\left( 0\right) \cdot \mathbf{p}_{\Vert }\left( 0\right) +\frac{3}{4}%
p_{z}\left( 0\right) p_{z}\left( 0\right) \right) .  \notag
\end{eqnarray}%
The Hartree-Fock solution for the order parameters is 
\begin{eqnarray}
p_{z}\left( 0\right) &=&\frac{1}{2}-\frac{\overline{\Delta }_{B}}{\overline{%
\Delta }_{B}^{\left( 1\right) }},  \label{symm} \\
\left\vert \mathbf{p}_{\Vert }\left( 0\right) \right\vert &=&\sqrt{\frac{1}{4%
}-\left\langle \rho _{z}\left( 0\right) \right\rangle ^{2}}.  \notag
\end{eqnarray}%
At the critical field 
\begin{equation}
\overline{\Delta }_{B}^{\left( 1\right) }=\frac{1}{4\beta }\sqrt{\frac{\pi }{%
2}},  \label{existence}
\end{equation}%
all electrons are pushed in level $n=1$ and the orbital coherence is lost.
For $B=10$ T, we find that $\overline{\Delta }_{B}^{\left( 1\right) }=3.536.$
From Eq. (\ref{symm}), we have the symmetry%
\begin{eqnarray}
\left\vert \mathbf{p}_{\Vert }\left( 0\right) \right\vert _{\Delta _{B}}
&=&\left\vert \mathbf{p}_{\Vert }\left( 0\right) \right\vert _{\Delta
_{B}^{\left( 1\right) }-\Delta _{B}}, \\
p_{z}\left( 0\right) _{\Delta _{B}} &=&-p_{z}\left( 0\right) _{\Delta
_{B}^{\left( 1\right) }-\Delta _{B}}.
\end{eqnarray}

The band structure consists of two dispersionless bands with energy 
\begin{eqnarray}
E_{+} &=&-\frac{1}{2}\sqrt{\frac{\pi }{2}}\left( \frac{e^{2}}{\kappa \ell }%
\right) , \\
E_{-} &=&-\sqrt{\frac{\pi }{2}}\left( \frac{e^{2}}{\kappa \ell }\right) .
\end{eqnarray}%
The energy in the middle of these two bands is 
\begin{equation}
\Delta _{B}^{(\ast )}=\frac{1}{2}\left( E_{+}+E_{-}\right) =-\frac{3}{4}%
\sqrt{\frac{\pi }{2}}\left( \frac{e^{2}}{\kappa \ell }\right) .
\label{deltaetoile}
\end{equation}%
We obviously have the symmetry%
\begin{equation}
E_{+}=2\Delta _{B}^{(\ast )}-E_{-}.
\end{equation}

\begin{figure}[tbph]
\includegraphics[scale=1]{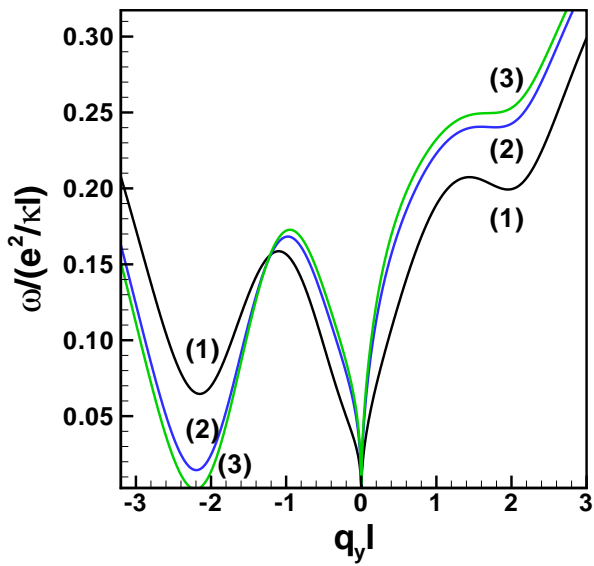}
\caption{(Color online) Dispersion relation of the orbital pseudospin mode
in the uniform phase for different values of the bias. For (1) to (3)
respectively: $\Delta _{B}/\left( e^{2}/\protect\kappa \ell \right)
=0.1,0.3,0.37.$}
\label{fig_instabilite}
\end{figure}

As can be seen from Eq. (\ref{liquidh}), the energy of the UP\ is
independent of the orientation of $\mathbf{p}_{\Vert }\left( 0\right) $ in
the $x-y$ plane. The UP is a quantum Hall orbital pseudomagnet. It follows
that this phase supports a gapless $x-y$ orbital pseudospin wave excitation,
a Goldstone mode related to the breaking of the rotation symmetry about the $%
z$ axis. The pseudospins remain parallel during their motion in this mode
(for $\mathbf{q}=0$) and so the DM\ term does not open a gap as it does for
the corresponding mode in the helical state (see below). The contribution of
the DM is strongest at larger bias and shorter wavelength. The mode's
dispersion relation is plotted in Fig. \ref{fig_instabilite} for different
values of the bias. It is highly anisotropic\cite{CoteOrbital} with an
unusual square root dispersion in the $q_{y}$ direction if the dipoles are
oriented in the direction $-\widehat{\mathbf{x}}$. The DM\ instability of
the UP occurs at $\overline{\Delta }_{B}=0.37$ (as well as at $\Delta
_{B}^{\left( 1\right) }-0.37$ if the bias is decreased from $\Delta
_{B}^{\left( 1\right) }$) and at finite wave vector $q_{y}\ell \approx -2.$
This suggests a transition to a ground state where the orbital pseudospin
field is no longer uniform as in the HP. In reality, we find that this
instability is preempted by a transition to a Skyrmion crystal at $\overline{%
\Delta }_{B}=0.06$ as we discussed in the previous section.

\subsection{Helical phase}

The exact helical phase (HP)\ given by the solution of the Hartree-Fock\
equations is too complex to study analytically and so we give the numerical
solution below. However it is instructive, in order to understand the
instability of the UP, to look at the energy of a simple helical phase (SHP)
with only one wave vector: 
\begin{equation}
\mathbf{p}_{SHP}\left( \mathbf{r}\right) =\frac{\alpha }{2S}p_{0}\left[ 
\mathbf{e}_{1}\cos \left( \mathbf{q}\cdot \mathbf{r}\right) +\mathbf{e}%
_{2}\sin \left( \mathbf{q}\cdot \mathbf{r}\right) \right] +\frac{\alpha }{2S}%
\eta \widehat{\mathbf{z}}\mathbf{,}  \label{exacthelix}
\end{equation}%
where $\mathbf{e}_{1},\mathbf{e}_{2}$ are orthogonal unit vectors and $%
\alpha $ is a normalization factor. The last term with $\eta $ describes the
uniform polarization due to the bias. It makes the magnitude of $\mathbf{p}%
_{SHP}$ position-dependent. It is however possible to impose an average
normalization condition. For simplicity, we will choose bias $\Delta
_{B}^{(1)}/2$ where $\beta =0\mathbf{\ }$and $\alpha =1.$ We define the unit
vectors:%
\begin{eqnarray}
\mathbf{e}_{1} &=&\sin \left( \theta _{1}\right) \cos \left( \varphi
_{1}\right) \widehat{\mathbf{x}}  \label{dingo} \\
&&+\sin \left( \theta _{1}\right) \sin \left( \varphi _{1}\right) \widehat{%
\mathbf{y}}+\cos \left( \theta _{1}\right) \widehat{\mathbf{z}},  \notag \\
\mathbf{e}_{2} &=&\sin \left( \theta _{2}\right) \cos \left( \varphi
_{2}\right) \widehat{\mathbf{x}}  \notag \\
&&+\sin \left( \theta _{2}\right) \sin \left( \varphi _{2}\right) \widehat{%
\mathbf{y}}+\cos \left( \theta _{2}\right) \widehat{\mathbf{z}},  \notag \\
\widehat{\mathbf{q}} &=&\cos \left( \varphi _{\mathbf{q}}\right) \widehat{%
\mathbf{x}}+\sin \left( \varphi _{\mathbf{q}}\right) \widehat{\mathbf{y}}. 
\notag
\end{eqnarray}%
The energy of the SHP is then%
\begin{eqnarray}
&&\frac{E_{SHP}}{N}/\left( \frac{e^{2}}{\kappa \ell }\right)  \label{eteta}
\\
&=&-\frac{11}{32}\sqrt{\frac{\pi }{2}}-\frac{1}{2}\beta \overline{\Delta }%
_{B}  \notag \\
&&+\frac{1}{16}a\left( q\right) \left[ \sin ^{2}\left( \theta _{1}\right)
+\sin ^{2}\left( \theta _{2}\right) \right]  \notag \\
&&+\frac{1}{16}b\left( q\right) \sin ^{2}\left( \theta _{1}\right) \cos
\left( 2\varphi _{1}-2\varphi _{\mathbf{q}}\right)  \notag \\
&&+\frac{1}{16}b\left( q\right) \sin ^{2}\left( \theta _{2}\right) \cos
\left( 2\varphi _{2}-2\varphi _{\mathbf{q}}\right)  \notag \\
&&+\frac{1}{16}c\left( q\right) \left[ \cos ^{2}\left( \theta _{1}\right)
+\cos ^{2}\left( \theta _{2}\right) \right]  \notag \\
&&-\frac{1}{16}d\left( q\right) \sin \left( \theta _{1}\right) \cos \left(
\theta _{2}\right) \cos \left( \varphi _{1}-\varphi _{\mathbf{q}}\right) 
\notag \\
&&+\frac{1}{16}d\left( q\right) \sin \left( \theta _{2}\right) \cos \left(
\theta _{1}\right) \cos \left( \varphi _{2}-\varphi _{\mathbf{q}}\right) . 
\notag
\end{eqnarray}%
In this equation, $\theta _{1},\varphi _{1}$ and $\theta _{2},\varphi _{2}$
must be choosen to make $\mathbf{e}_{1}\cdot \mathbf{e}_{2}=0.$ For values
of $q$ where all interactions $a\left( q\right) ,...,d\left( q\right) $ are
negative, inspection of Eq. (\ref{eteta}) and Fig. \ref{interactions} shows
that the lowest-energy solution is obtained when $\theta _{1}=0,\theta
_{2}=\pi /2$ and $\varphi _{2}=\varphi _{\mathbf{q}}.$ The vector $\mathbf{e}%
_{2}$ is free to take any orientation in the $x-y$ plane. The plane of
polarization of the helix is thus $\widehat{\mathbf{z}}-\mathbf{e}_{2}$ and
its energy is given by 
\begin{eqnarray}
\frac{E_{SHP}}{N}/\left( \frac{e^{2}}{\kappa \ell }\right) &=&-\frac{11}{32}%
\sqrt{\frac{\pi }{2}}-\frac{1}{2}\beta \overline{\Delta }_{B} \\
&&+\frac{1}{16}\left( a\left( q\right) +b\left( q\right) +c\left( q\right)
+d\left( q\right) \right) .  \notag
\end{eqnarray}%
The sum of the four interactions is plotted in Fig. \ref{interactions}. From
this figure, we see that the wave vector that minimizes the energy of the
helix is $q_{0}\ell \approx 2.3.$ Note that an helix in the $x-y$ plane
(with $\theta _{1}=\theta _{2}=\pi /2,\varphi _{1}=0,\varphi _{2}=\pi /2$)
has a higher energy given by 
\begin{equation}
\frac{E_{SHP}^{\prime }}{N}/\left( \frac{e^{2}}{\kappa \ell }\right) =-\frac{%
11}{32}\sqrt{\frac{\pi }{2}}-\frac{1}{2}\beta \overline{\Delta }_{B}+\frac{1%
}{16}a\left( q\right) .
\end{equation}

Fig. \ref{interactions} shows that the Heisenberg exchange and dipolar parts
of the energy of the helix (the curve labelled $a+b+c$) has its minimum at $%
q=0.$ We conclude that these interactions do not favor the formation of a
non-uniform state. In our system, the DM\ interaction is responsible for the
formation of the helix with a finite wavevector. We remark that our helix
has its wave vector $\mathbf{q}$ parallel to $\mathbf{e}_{2}$ instead of $%
\widehat{\mathbf{n}}=\mathbf{e}_{1}\times \mathbf{e}_{2}$ as is often the
case.

\subsubsection{Energy}

When we solve for the exact helical phase using Eq. (\ref{EHF}), we find a
multicomponent helix with a finite value of $\eta _{z}$. We choose $\mathbf{e%
}_{2}=\widehat{\mathbf{x}}$ so that the helix wave vector $\mathbf{q}=nq_{0}%
\widehat{\mathbf{x}}$ with $q_{0}=2\pi /\lambda $ and the pseudospin vector $%
\left\langle \overrightarrow{\rho }\right\rangle $ rotates clockwise in the $%
z-x$ plane. We find that the optimal period of the helix is $\lambda /\ell
=2.74+0.01(\overline{\Delta }_{B}-\overline{\Delta }_{B}^{\left( 1\right)
}/2)^{2},$ close to the value we found above for the SHP, and varies only
slightly with bias. Since $\ell \left( \text{\AA }\right) =256/\sqrt{B\left( 
\text{T}\right) },$ the period of the helix is of the order $220$ \AA\ for $%
B=10$ T.

\begin{figure}[tbph]
\includegraphics[scale=1]{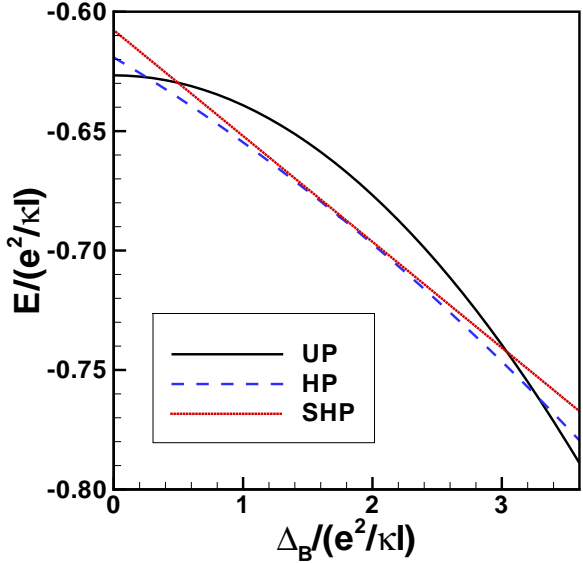}
\caption{(Color online) Energy of uniform (UP), exact helix (HP), and simple
helix (SHP) as a function of bias at $B=10$ T.}
\label{energie}
\end{figure}

The energy of the numerical solution is independent of the direction of the
wave vector $\mathbf{q}$ in the $x-y$ plane, as expected. As for the SHP,
the plane of polarization is given by $\left( \mathbf{e}_{1},\mathbf{e}%
_{2}\right) =\left( \mathbf{q,}\widehat{\mathbf{z}}\right) \mathbf{.}$The
energy of the HP is plotted in Fig. \ref{energie} along with that of the
SHP. The energy of the HP is obviously very close to that of the SHP. We
find indeed that the number of Fourier components in the HP is small.

\subsubsection{Peudospin pattern in real space}

\begin{figure}[tbph]
\includegraphics[scale=1]{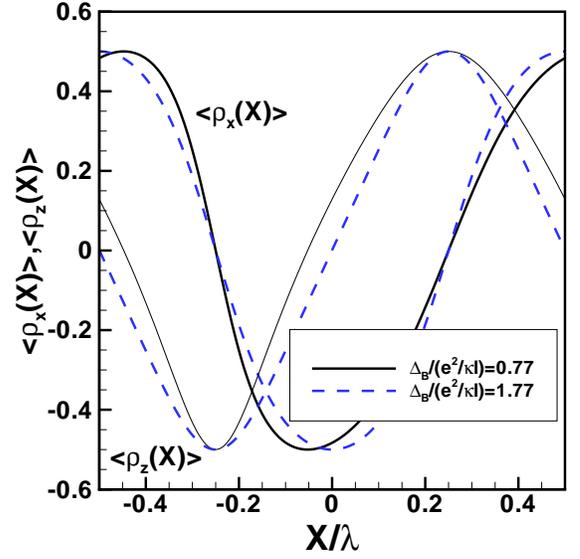}
\caption{(Color online) Density profile for the pseudospin fields $%
\left\langle \protect\rho _{x}\left( X\right) \right\rangle $ and $%
\left\langle \protect\rho _{z}\left( X\right) \right\rangle $ in the helical
phase for two different biases. }
\label{sxsz}
\end{figure}

Fig. \ref{sxsz} shows the orbital pseudospin field in real space at
different biases. The following symmetry relations hold%
\begin{eqnarray}
\left\langle \rho _{x}\left( x\right) \right\rangle _{\Delta _{B}}
&=&\left\langle \rho _{x}\left( -x\right) \right\rangle _{\Delta
_{B}^{(1)}-\Delta _{B}}, \\
\left\langle \rho _{z}\left( x\right) \right\rangle _{\Delta _{B}}
&=&-\left\langle \rho _{z}\left( -x\right) \right\rangle _{\Delta
_{B}^{(1)}-\Delta _{B}}.
\end{eqnarray}%
In the numerical solution, $\left\langle \rho _{x}\left( \mathbf{q}=0\right)
\right\rangle =\left\langle \rho _{y}\left( \mathbf{q}=0\right)
\right\rangle =0$ but $\left\langle \rho _{z}\left( \mathbf{q}=0\right)
\right\rangle \neq 0$ (except at the special bias $\overline{\Delta }%
_{B}^{\left( 1\right) }/2$). As expected, the helix has a finite value for
the term $\mathbf{\eta }$ in Eq. (\ref{exacthelix}) because the bias tilts
the pseudospin vector away from the $z$ axis$.$

In the helical phase $\left\langle \rho _{0,0}\left( \mathbf{q}\right)
\right\rangle +\left\langle \rho _{1,1}\left( \mathbf{q}\right)
\right\rangle =1$ and so the orbital density $\left\langle \rho _{n}\left( 
\mathbf{r}\right) \right\rangle =1/2\pi \ell ^{2}$ is constant in real
space. The densities%
\begin{eqnarray}
\left\langle \rho _{0,0}\left( \mathbf{r}\right) \right\rangle &=&\frac{1}{2}%
+\left\langle \rho _{z}\left( \mathbf{r}\right) \right\rangle ,  \label{ro00}
\\
\left\langle \rho _{1,1}\left( \mathbf{r}\right) \right\rangle &=&\frac{1}{2}%
-\left\langle \rho _{z}\left( \mathbf{r}\right) \right\rangle .  \label{ro11}
\end{eqnarray}

\subsubsection{Band structure}

For a modulation of the pseudospin density in only one direction such as in
the helical state, Eq. (\ref{EHF}) for the single-particle Green's function
can be written more simply as%
\begin{equation}
i\omega _{n}G_{n,m}\left( X,\omega _{n}\right) -\sum_{k}\Lambda _{n,k}\left(
X\right) G_{k,m}\left( X,\omega _{n}\right) =\delta _{n,m},  \label{gf}
\end{equation}%
with $n,m,k=0,1$ where%
\begin{equation}
G_{n,m}\left( X,\tau \right) =-\left\langle T_{\tau }c_{n,X}\left( \tau
\right) c_{m,X}^{\dagger }\left( 0\right) \right\rangle ,
\end{equation}%
and%
\begin{eqnarray}
\Lambda _{n,m}\left( X\right) &=&-\beta \Delta _{B}\delta _{n,m}\delta _{n,1}
\\
&&+\sum_{q}\left( U^{H}\left( n,m,q\right) -U^{F}\left( n,m,q\right) \right)
e^{iqX}.  \notag
\end{eqnarray}%
The order parameters are given by

\begin{equation}
\left\langle \rho _{n,m}\left( \mathbf{q}\right) \right\rangle =\frac{1}{%
N_{\varphi }}\delta _{q_{y},0}\sum_{X}e^{-iq_{x}X}\left\langle \rho
_{n,m}\left( X\right) \right\rangle ,
\end{equation}%
with%
\begin{equation}
\left\langle \rho _{n,m}\left( X\right) \right\rangle =G_{m,n}\left( X,\tau
=0^{-}\right) .
\end{equation}

We can solve formally for the Green's function to find 
\begin{eqnarray}
G_{n,m}\left( X,i\omega _{n}\right) &=&\frac{A_{n,m}\left( X\right) }{%
i\omega _{n}+\mu /\hslash -E_{+}\left( X\right) /\hslash } \\
&&+\frac{B_{n,m}\left( X\right) }{i\omega _{n}+\mu /\hslash -E_{-}\left(
X\right) /\hslash }.  \notag
\end{eqnarray}%
The band structure consists of two bands with dispersion $E_{\pm }\left(
X\right) $ given by 
\begin{eqnarray}
E_{\pm }\left( X\right) &=&\frac{1}{2}\left[ \Lambda _{0,0}\left( X\right)
+\Lambda _{1,1}\left( X\right) \right] \\
&&\pm \frac{1}{2}\sqrt{\left[ \Lambda _{0,0}\left( X\right) -\Lambda
_{1,1}\left( X\right) \right] ^{2}+4\left\vert \Lambda _{0,1}\left( X\right)
\right\vert ^{2}},  \notag
\end{eqnarray}%
with%
\begin{eqnarray}
A_{n,m}\left( X\right) &=&\frac{E_{+}\left( X\right) \delta _{n,m}-\Lambda _{%
\overline{m},\overline{n}}\left( X\right) }{E_{+}\left( X\right)
-E_{-}\left( X\right) },  \label{h1} \\
B_{n,m}\left( X\right) &=&\frac{\Lambda _{\overline{m},\overline{n}}\left(
X\right) -E_{-}\left( X\right) \delta _{n,m}}{E_{+}\left( X\right)
-E_{-}\left( X\right) },  \notag
\end{eqnarray}%
and $\overline{n}=1-n$ , etc. At $T=0$ K, 
\begin{equation}
\left\langle \rho _{n,m}\left( X\right) \right\rangle =B_{m,n}\left(
X\right) .  \label{h2}
\end{equation}%
It is easy to show from Eq. (\ref{gf}) that, at $T=0$ K, we have the sum
rule 
\begin{equation}
\left\vert \left\langle \overrightarrow{\rho }\left( X\right) \right\rangle
\right\vert ^{2}=\frac{1}{4}.  \label{rulep}
\end{equation}%
Moreover, From Eqs. (\ref{h1},\ref{h2}), we have also%
\begin{equation}
\left\langle \rho _{0,0}\left( X\right) \right\rangle +\left\langle \rho
_{1,1}\left( X\right) \right\rangle =1,
\end{equation}%
so that the total guiding-center density is unmodulated in the spiral phase.
Note that the real density $n\left( \mathbf{r}\right) $ given by Eq. (\ref%
{realdensity}) is modulated however. The modulus of the pseudospin vector $%
\left\vert \left\langle \overrightarrow{\rho }\left( X\right) \right\rangle
\right\vert =1/2$ is constant in space. This is not the case when there is a
two-dimensional modulation of the pseudospin texture as we will see in the
crystal phase.

\begin{figure}[tbph]
\includegraphics[scale=1]{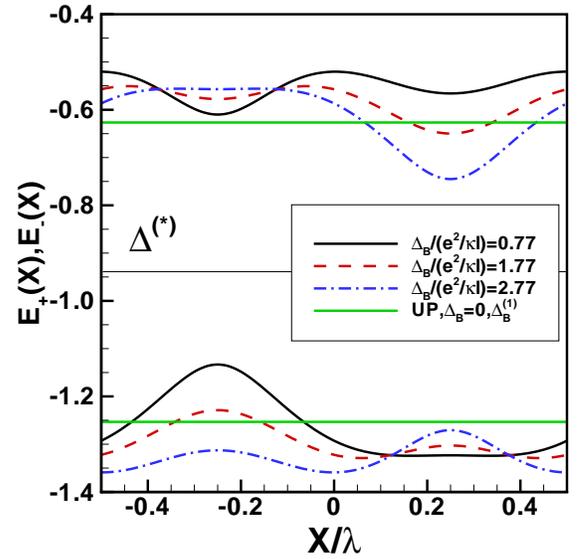}
\caption{(Color online) Band structure $E_{\pm }\left( X\right) $ in the
helical state for different biases and for the uniform phase (straight
lines) at $\Delta _{B}=0$ or $\Delta _{B}=\Delta _{B}^{(1)}.$ }
\label{band}
\end{figure}

The band structure of the HP is shown in Fig. \ref{band} for different
values of the bias. The lowest band $E_{-}\left( X\right) $ is completely
filled so that the system is insulating. There is a continuum of
electron-hole excitations in the energy range $E_{eh}\in \left[ 0.56,0.80%
\right] \left( e^{2}/\kappa \ell \right) $ which is roughly independent of
the bias. The band structure has the symmetry 
\begin{equation}
\left. E_{+}\left( X\right) \right\vert _{\Delta _{B}}=2\Delta _{B}^{(\ast
)}-\left. E_{-}\left( -X\right) \right\vert _{\Delta _{B}^{(1)}-\Delta _{B}},
\end{equation}%
where $\Delta _{B}^{(\ast )}$ was defined previously in Eq. (\ref%
{deltaetoile}).

\subsubsection{Density of states}

\begin{figure}[tbph]
\includegraphics[scale=1]{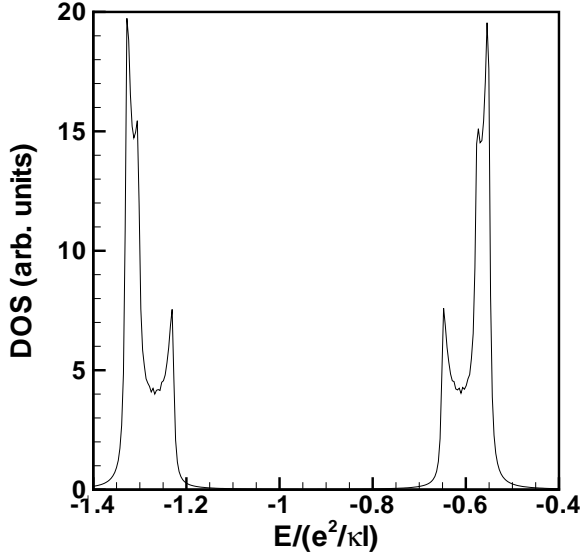}
\caption{Density of states in the helical state at bias $\Delta _{B}=\Delta
_{B}^{(1)}/2.$ }
\label{dos}
\end{figure}

We compute the density of states (DOS) from the retarded single-particle
Green's function i.e.%
\begin{eqnarray}
g\left( E,\Delta _{B}\right) &=&-\frac{N_{\varphi }}{\pi }\sum_{n}\Im \left[
G_{n,n}^{R}\left( \mathbf{q}=0,E\right) \right] , \\
&=&-\frac{1}{\pi }\sum_{n}\sum_{X}\Im \left[ G_{n,n}^{R}\left( X,X,E\right) %
\right] .  \notag
\end{eqnarray}%
The DOS\ is represented in Fig. \ref{dos} for $\overline{\Delta }_{B}=\Delta
_{B}^{(1)}/2.$ At this bias, the lower energy band is the mirror image (with
respect to a mirror line at the energy $\Delta _{B}^{(\ast )}$) of the high
energy band. More generally, because of the symmetry of the band stucture,
the DOS has the corresponding symmetry%
\begin{equation}
\left. g\left( E\right) \right\vert _{\Delta _{B}}=\left. g\left( 2\Delta
_{B}^{(\ast )}-E\right) \right\vert _{\Delta _{B}^{(1)}-\Delta _{B}}.
\end{equation}%
The extrema in the band structure shown in Fig. \ref{band} lead to
distinctive van-Hove singularities in the DOS as seen in Fig. \ref{dos}.

\subsubsection{Response functions and collective modes}

We compute numerically the 16 retarded response functions in the GRPA\ using
Eq. (\ref{grpa1}). In the non-uniform phases, one must calculate $\chi
_{i,i,j,j}^{R}\left( \mathbf{k+G},\mathbf{k+G},\omega \right) $ for all
recriprocal lattice vector $\mathbf{G}$. Using the transformations given by
Eq. (\ref{pseudospin}), we obtain the response function for the pseudospin
operators i.e. $\chi _{a,b}^{R}$ with $a,b=\rho _{n},\rho _{x},\rho
_{y},\rho _{z}.$

\begin{figure}[tbph]
\includegraphics[scale=1]{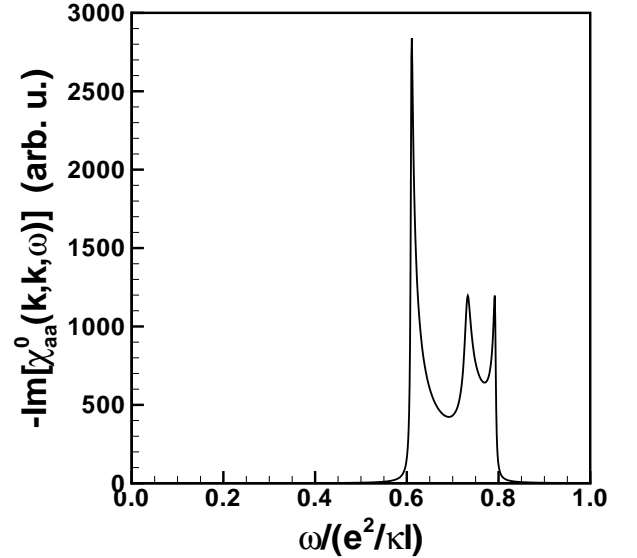}
\caption{Imaginary part of the density response function ($a=\protect\rho %
_{n}$) evaluated in the Hartree-Fock approximation (single-bubble
approximation) in the helical phase for $\mathbf{k}=\left( 0,1\right) 2%
\protect\pi /\protect\lambda $ and $\Delta _{B}/\left( e^{2}/\protect\kappa %
\ell \right) =1.0.$}
\label{chizero}
\end{figure}

We show in Fig. \ref{chizero} the imaginary part of the response function $%
\chi _{\rho _{n},\rho _{n}}^{0,R}\left( \mathbf{k},\mathbf{k},\omega \right) 
$ defined in Eq. (\ref{grpa2}). This function corresponds to the
single-bubble approximation and does not capture the collective modes but
only the particle-hole continuum. The continuum appears in the range $%
E_{eh}\in \left[ 0.56,0.80\right] $ in accordance with the band structure
calculation.

\begin{figure}[tbph]
\includegraphics[scale=1]{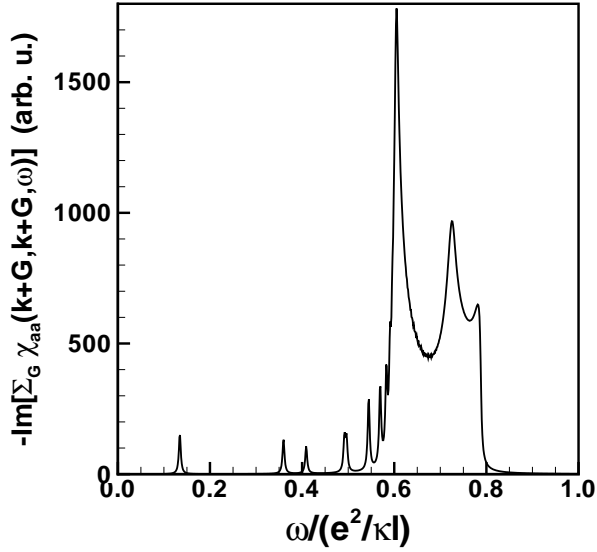}
\caption{Imaginary part of the density response function ($a=\protect\rho %
_{n}$) evaluated in the GRPA in the helical phase for $\mathbf{k}=\left(
0,1\right) 2\protect\pi /\protect\lambda $ and $\Delta _{B}/\left(
e^{2}/\left( \ell \right) \right) =1.0.$}
\label{chigrpa}
\end{figure}

The collective modes can be obtained from the poles of the imaginary part of
the full GRPA$\ $response functions $\chi _{a,a}^{R}\left( \mathbf{k},%
\mathbf{k},\omega \right) .$ To get the dispersion relation, we follow the
frequencies of these poles as the wave vector is varied in the Brillouin
zone. We remark that, in order to capture the electron-hole continuum from $%
\chi _{a,a}^{R}$ computed in the GRPA, we must sum over all the reciprocal
lattice vectors i.e. compute 
\begin{equation}
\chi _{a,a}^{R}\left( \mathbf{k},\omega \right) \equiv \sum_{\mathbf{G}}\chi
_{a,a}^{R}\left( \mathbf{k+G},\mathbf{k+G},\omega \right) .  \label{chi}
\end{equation}%
This function is shown in Fig. \ref{chigrpa} where the electron-hole
continuum together with the collective modes are clearly visible. Note that
all response functions are coupled in the GRPA\ equations. Consequently,
they all share the same poles. However, the weight of a given pole depends
on the nature of the underlying mode and is not the same in all response
functions. Electron-hole excitations appear as very localized excitations
and are captured in the response functions at finite $\mathbf{G}.$

\begin{figure}[tbph]
\includegraphics[scale=1]{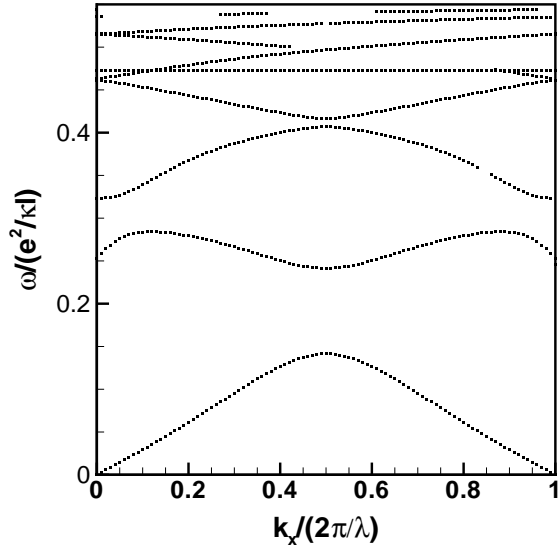}
\caption{Dispersion relation of the first collective modes of the helical
phase along $k_{x}$ for $\Delta _{B}/(e^{2}/\protect\kappa \ell )=1.0.$ The
rotation of the pseudospins is along the $\widehat{\mathbf{x}}$ direction
with a period $\protect\lambda .$}
\label{dispersionkx}
\end{figure}

\begin{figure}[tbph]
\includegraphics[scale=1]{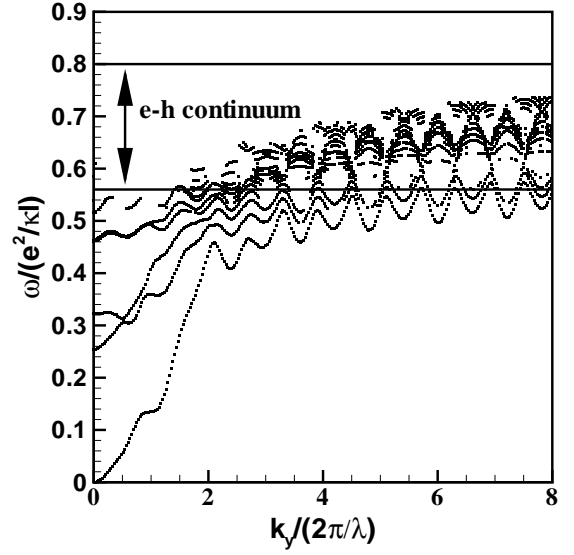}
\caption{Dispersion relation of the collective modes of the helical phase
along $k_{y}$ for $\Delta _{B}/(e^{2}/\protect\kappa \ell )=1.0.$ The
rotation of the pseudospins are along the $\widehat{\mathbf{x}}$ direction
with a period $\protect\lambda .$The two lines indicates the region of the
electron-hole continuum where the collective modes are damped.}
\label{dispersionky}
\end{figure}

Fig. \ref{dispersionkx} shows the dispersion of the first collective modes
of the helical phase for $\overline{\Delta }_{B}=1.0$ along the direction of
the pseudospin modulation i.e. along $k_{x}$. (The absence of points in the
dispersion of some of the modes is a numerical artefact.)\ The real-space
pattern is periodic with period $\lambda $ along $k_{x}$ but there is no
periodicity in the dispersion in the $k_{y}$ direction as shown in Fig. \ref%
{dispersionky}. We have indicated in Fig. \ref{dispersionky} the region of
the electron-hole continuum where the collective modes are damped. The
higher-energy collective modes are less dispersive and correspond to more
localized excitations which eventually vanish in the electron-hole
continuum. Since we are dealing with a continuous structure, the number of
collective modes is not finite. This is also true in the Skyrmion crystal
phase that we study in the next section.

The unusual oscillations in the dispersion relation in the $k_{y}$ direction
are due to the phase factor $\gamma _{\mathbf{q},\mathbf{q}^{\prime }}=\exp
\left( -i\mathbf{q}\times \mathbf{q}^{\prime }\ell ^{2}/2\right) =\exp
\left( -i\left( \mathbf{k}+\mathbf{G}\right) \times \left( \mathbf{k}+%
\mathbf{G}^{\prime }\right) \ell ^{2}/2\right) $ in Eq. (\ref{grpa2}). In
the spiral phase, $\mathbf{G}=\left( 2\pi n/\lambda \right) \widehat{\mathbf{%
x}}$ and $\mathbf{k}$ is a vector in the first Brillouin zone. It follows
that $\gamma _{\mathbf{q},\mathbf{q}^{\prime }}=\exp \left( -ik_{y}\left(
G_{x}-G_{x}^{\prime }\right) \ell ^{2}/2\right) .$ In the direction $%
k_{y}=0, $ the phase factor $\gamma _{\mathbf{q},\mathbf{q}^{\prime }}$ is
constant and there is no oscillation. Along $k_{y}$, however, the phase
factor is periodic with periods $\lambda ^{2}/m\pi \ell ^{2}$ in units of $%
2\pi /\lambda .$ In Fig. \ref{dispersionky}, we can clearly see the periods
with $m=1,2,4$ using $\lambda /\ell \approx 2.74.$

The gapless mode of the helical phase is a phonon mode corresponding to a
global translation of the density $n\left( \mathbf{r}\right) $ in a space
(for $\mathbf{q}=0$) accompanied by an in-phase $z-x$ rotation of the all
the pseudospins. As can be seen from Eq. (\ref{EPP}), this is a global
symmetry of the Hamiltonian. At small wave vector $\mathbf{k}$, the
dispersion of the phonon mode has $\omega \left( k_{x},k_{y}=0\right) \sim
k_{x}$ (the wave vector of the helix is $\mathbf{q=}q_{0}\widehat{\mathbf{x}}
$) while $\omega \left( k_{x}=0,k_{y}\right) \sim k_{y}^{2.5}$.

The gapless $x-y$ pseudospin mode of the uniform phase acquires a gap in the
helical state. Since the bias fixes the $z$ component of the pseudospin in
the uniform phase, this mode corresponds to an oscillation of the
pseudospins at $\rho _{z}\left( \mathbf{r}\right) $ constant when $\mathbf{k}%
=0.$ In the helical phase, the mode that corresponds to this motion is
gapped because the DM\ term sets a preferred plane of rotation for the
pseudospins i.e. the direction of $\widehat{\mathbf{n}}$ and $\mathbf{q}$ in
Eq. (\ref{dingo}) are related. We remark that in a double quantum well
systems where a stripe state occurs at $\nu =1$ in higher Landau levels,
both the phonon and pseudospin modes are gapless in the absence of tunneling%
\cite{bouchiha}. There is no DM term in the Hamiltonian of the stripe state
and both motions correspond to a symmetry of the Hamiltonian. A plot (not
shown) of the susceptibilities $\chi _{a,a}^{R}\left( \mathbf{k\rightarrow 0}%
,\omega \right) $ shows that the lowest-energy gapped mode in Fig. \ref%
{dispersionky} has a weight in only $\chi _{\rho _{x},\rho _{x}}^{R}$ and $%
\chi _{\rho _{y},\rho _{y}}^{R}.$ It thus seems likely that this mode is the
gapped pseudospin $x-y$ mode.

\subsection{Skyrmion crystal phase}

\begin{figure}[tbph]
\includegraphics[scale=1]{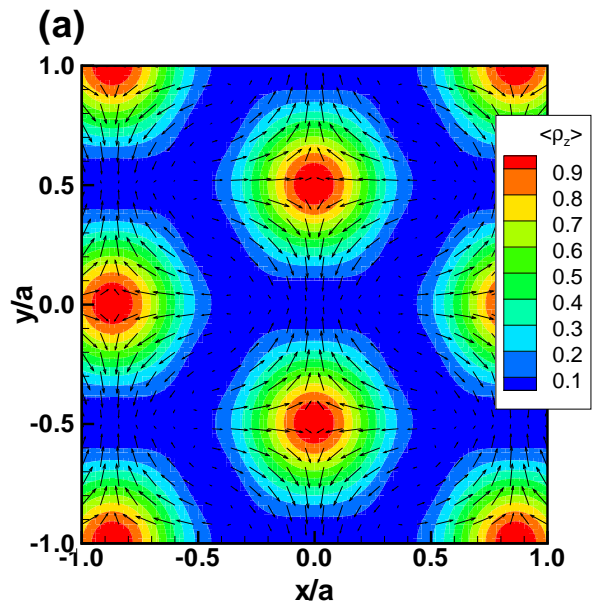}
\caption{(Color online)\ Pseudospin pattern $\left\langle \protect\rho %
_{x}\left( r\right) \right\rangle ,\left\langle \protect\rho _{y}\left(
r\right) \right\rangle $ in the Skyrmion crystal at bias $\Delta _{B}/\left(
e^{2}/\protect\kappa \ell \right) =0.2.$}
\label{cristal1}
\end{figure}

The skyrmion crystal phase occurs on both sides of the helical state in the
phase diagram. Fig. \ref{cristal1} shows the pseudospin fields defined in
the guiding-center representation by Eq. (\ref{pseudospin}) at bias $%
\overline{\Delta }_{B}=0.2.$ From Eq. (\ref{dipole}), the parallel component
of $\left\langle \overrightarrow{\rho }\left( \mathbf{r}\right)
\right\rangle $ is directly related to the physical electric dipoles. The
crystal at $\Delta _{B}^{(1)}-\Delta _{B}$ is the electron-hole conjugate of
that at $\Delta _{B}$ i.e.%
\begin{eqnarray}
\left. n\left( \mathbf{r}\right) \right\vert _{\Delta _{B}^{(1)}-\Delta
_{B}} &=&2-\left. n\left( \mathbf{r}\right) \right\vert _{\Delta _{B}},
\label{sym1} \\
\left. \left\langle \rho _{z}\left( \mathbf{r}\right) \right\rangle
\right\vert _{\Delta _{B}^{(1)}-\Delta _{B}} &=&-\left. \left\langle \rho
_{z}\left( \mathbf{r}\right) \right\rangle \right\vert _{\Delta _{B}}.
\end{eqnarray}%
The pseudospin vorticity in the $x-y$ plane is however the same for both
crystals. As in the helical state, the electron-hole conjugation applies to
the other properties described in this section.

In the crystal phase, the pseudospin $\left\langle \overrightarrow{\rho }%
\left( \mathbf{r}\right) \right\rangle $ is not constant in space. In fact, $%
\left\langle \overrightarrow{\rho }\left( \mathbf{r}\right) \right\rangle $
is a pseudospin \textit{density} not a unit field. In Fig. \ref{cristal1},
the pseudospin density has not been normalized and so $\left\langle \rho
_{z}\left( \mathbf{r}\right) \right\rangle $ is greater than $1/2$ in some
regions of the crystal. The guiding-center density $\left\langle \rho \left( 
\mathbf{r}\right) \right\rangle =1$ is however constant. The density in each
orbital is given by Eqs. (\ref{ro00},\ref{ro11}). The difference between the
orbital and crystal cases is that $\left\langle \rho _{z}\left( \mathbf{r}%
\right) \right\rangle \in \left[ -1/2,1/2\right] $ in the orbital state
while $\left\langle \rho _{z}\left( \mathbf{r}\right) \right\rangle $ is
positive for $\Delta _{B}<\Delta _{B}^{(1)}/2$ and negative for $\Delta
_{B}>\Delta _{B}^{(1)}/2$ in the crystal phase, in the regions where the
crystal is the ground state.

The crystal state is constructed by assuming that the number of vortices is
equal to the number of electrons. Since the number of flux quanta is equal
to the number of electrons at filling factor $\nu =1$, we have $\nu =2\pi
n\ell ^{2}$ with $n=1/\varepsilon a^{2}$ where $\varepsilon =\sqrt{3}/2$ for
a triangular lattice and the lattice constant is given by $a/\ell =\sqrt{%
2\pi }/\varepsilon $ . The lattice spacing is thus controlled by the
magnetic field.

The pseudospin pattern at each crystal site in Fig. \ref{cristal1}
ressembles that of a meron (half a Skyrmion) with negative charge. The $z$
component of the pseudospin is up at the center and goes in the $x-y$ plane
in-between the crystal sites. The vorticity of the field $\left\langle 
\overrightarrow{\rho }_{\Vert }\left( \mathbf{r}\right) \right\rangle $ is
positive. Evaluation of the real density $n\left( \mathbf{r}\right) $ shows
that each meron is placed on a uniform background of negative charge $%
n\left( \mathbf{r}\right) $ i.e. the density does not go to zero between two
crystal sites. Approximatively half of the real density is in the uniform
background and half is in the merons. We could thus describe the state at $%
\Delta _{B}<\Delta _{B}^{(1)}/2$ as a crystal of charged merons on top of a
uniform background of electrons. For $\Delta _{B}>\Delta _{B}^{(1)}/2$, the
vorticity is still positive, but the $z$ component of the pseudospin is down
at the center of the vortices and we have positively charged anti-merons (or
holes in the electronic density). Note that the electronic density inside
each meron is not quantized in our approach since we work with a pseudospin
density and the pseudospin modulus changes with position. We could think of
the crystal state has a two-dimensional charge-density-wave with an
amplitude that can change continuously with bias. We use the term
\textquotedblleft Skyrmion crystal\textquotedblright\ in a loose way to
refer to that state.

\subsubsection{Density of states}

\begin{figure}[tbph]
\includegraphics[scale=1]{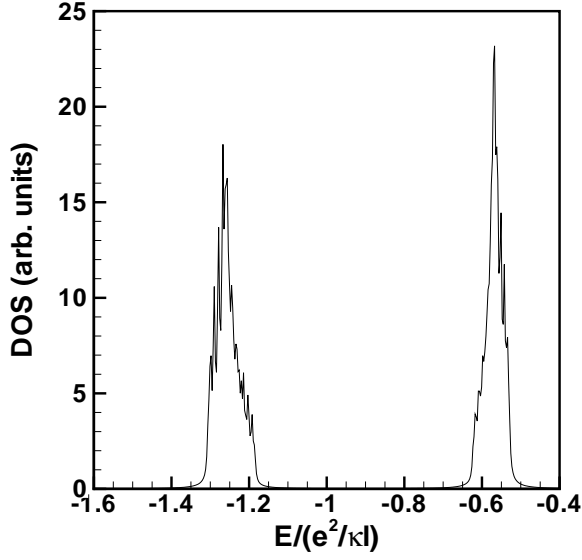}
\caption{Density of states of the Skyrmion crystal phase at bias $\Delta
_{B}/\left( e^{2}/\protect\kappa \ell \right) =0.2.$}
\label{doscristal}
\end{figure}

The density of state is shown in Fig. \ref{doscristal}. It is closer in
shape to the DOS\ of the helical phase than to that of a crystal as can be
seen by comparing Fig. \ref{doscristal} with the DOS\ of a Skyrmion crystal
at $\nu =1.2$ and $\overline{\Delta }_{B}=1.28$ shown in Fig. 13 of Ref. %
\onlinecite{CoteCrystal}.

\subsubsection{Collective modes}

\begin{figure}[tbph]
\includegraphics[scale=1]{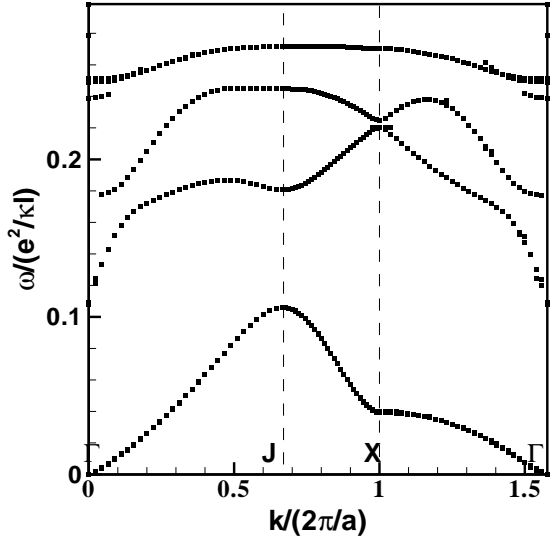}
\caption{Dispersion relation of the first low-energy collective modes of the
Skyrmion crystal at $\Delta _{B}/\left( e^{2}/\protect\kappa \ell \right)
=0.2$ along the path \ $\Gamma -J-X-\Gamma $ in the irreducible Brillouin
zone.}
\label{discristal}
\end{figure}

The dispersion relation of the collective modes of the skyrmion crystal is
shown in Fig. \ref{discristal}. Only the first low-energy modes are shown.
The modes become more dense at higher energy until the electron-hole
continuum is reached. From Fig. \ref{doscristal}, the continuum is in the
range $E_{eh}\in \left[ 0.47,0.90\right] .$ The dispersion for $\Delta
_{B}^{(1)}-\Delta _{B}$ (not shown) is exactly the same as that for $\Delta
_{B}$ as expected.

Since our calculation does not include disorder, the lowest-energy mode is a
gapless phonon mode as in the helical state. For $\overline{\Delta }_{B}=0.1$%
, the dispersion is isotropic with $\omega \sim k^{1.5}$ at small wave
vector as in a Wigner crystal. The pseudospin $x-y$ mode which is usually
gapless in a Skyrme crystal\cite{coteskyrme} is now gapped because of the DM
term in the pseudospin Hamiltonian. A plot (not shown) of the
susceptibilities $\chi _{a,a}^{R}\left( \mathbf{k\rightarrow 0},\omega
\right) $ shows that the lowest-energy gapped mode in Fig. \ref{discristal}
has substantial weights in $\chi _{\rho _{x},\rho _{x}}^{R}$ and $\chi
_{\rho _{y},\rho _{y}}^{R}.$ It thus seems likely that this mode is the
gapped pseudospin $x-y$ mode.

The energy difference between the crystal and helical phases is quite small,
of the order of a few percent. It may be then, that these phases will be
very sensitive to disorder as well as thermal and quantum fluctuations. An
evaluation of the effect of these perturbations is however beyond the scope
of the present paper.

\section{ELECTROMAGNETIC ABSORPTION}

The collective modes of the helical and Skyrmion crystal phases can be
detected in electromagnetic absorption experiments. With $\kappa =5$ for
graphene on SiO$_{2}$ substrate, we have $\nu _{0}=e^{2}/2\pi h\kappa \ell
=0.43\sqrt{B}$ THz with $B$ in Tesla. The frequency of the collective modes
at $\mathbf{q}=0$ in both the helical and crystal phases are in the range $%
\nu \in \left[ 0.1,0.6\right] \nu _{0}\approx \left[ 0.14,0.84\right] $ THz
for $B=10$ T.

Theoretically, the absorption can be related to the current-current
correlation function as explained in Ref. \onlinecite{CoteOrbital}. We give
here only the main results of this calculation.

We write the current operator, projected onto $N=0$ and valley $\mathbf{K}%
^{\prime }$ as 
\begin{eqnarray}
\mathbf{J} &=&-c\left. \frac{\partial H_{\mathbf{K}^{\prime }}^{0}}{\partial 
\mathbf{A}^{e}}\right\vert _{A_{i}^{e}=0} \\
&=&-\sqrt{2}\beta \Delta _{B}\frac{e\ell }{\hslash }N_{\varphi }\left( \rho
_{y}\left( 0\right) \widehat{\mathbf{x}}+\rho _{x}\left( 0\right) \widehat{%
\mathbf{y}}\right) ,  \notag
\end{eqnarray}%
where $\mathbf{A}^{e}$ is the vector potential of the external
electromagnetic field and $H_{\mathbf{K}^{\prime }}^{0}$ was defined in Sec.
II. We define the current-current correlation function Matsubara Green's
functions as 
\begin{eqnarray}
\chi _{J_{\alpha },J_{\beta }}\left( \tau \right) &=&-\frac{1}{A}%
\left\langle T_{\tau }J_{\alpha }\left( \tau \right) J_{\beta }\left(
0\right) \right\rangle \\
&=&\left( \frac{\Delta _{B}}{\gamma _{1}}\right) ^{2}\frac{e^{2}\hslash ^{2}%
}{4\pi m^{\ast 2}\ell ^{4}}\chi _{\rho _{\overline{\alpha }},\rho _{%
\overline{\beta }}}\left( \tau \right) ,  \notag
\end{eqnarray}%
where $A$ is the area of the 2DEG and 
\begin{equation}
\chi _{\rho _{\alpha },\rho _{\beta }}\left( \tau \right) =-\left\langle
T_{\tau }\rho _{\alpha }\left( 0,\tau \right) \rho _{\beta }\left(
0,0\right) \right\rangle
\end{equation}%
with $\alpha ,\beta =x,y.$ Note that $\overline{\alpha },\overline{\beta }$
are defined by $\overline{x}=y,\overline{y}=x.$

The electromagnetic absorption for an electric field oriented along the
direction $\alpha $ is given by%
\begin{eqnarray}
P_{\alpha }\left( \omega \right) &=&-\frac{1}{\hslash }\Im \left[ \frac{\chi
_{J_{\alpha },J_{\alpha }}^{R}\left( \omega \right) }{\omega +i\delta }%
\right] E_{0}^{2}  \label{mwabsorption} \\
&=&-\frac{1}{2}\left( \frac{e^{2}}{h}\right) \left( \frac{\Delta _{B}}{%
\gamma _{1}}\right) ^{2}\omega _{c}^{\ast 2}E_{0}^{2}  \notag \\
&&\times \Im \left[ \frac{\chi _{\rho _{\overline{\alpha }},\rho _{\overline{%
a}}}^{R}\left( \omega \right) }{\omega +i\delta }\right] ,  \notag
\end{eqnarray}%
where we have assumed a uniform electric field $\mathbf{E}\left( t\right)
=E_{0}\widehat{\mathbf{\alpha }}e^{-i\omega t}$ with polarization $\widehat{%
\mathbf{\alpha }}$ and taken the analytic continuation $i\Omega
_{n}\rightarrow \omega +i\delta $ of $\chi _{J_{\alpha },J_{\alpha }}\left(
0,i\Omega _{n}\right) $ to get the retarded response function. The response
functions $\chi _{\rho _{\alpha },\rho _{\beta }}^{R}\left( \omega \right) $
are calculated in units of $\hslash /\left( e^{2}/\kappa \ell \right) $ so
that $P_{\alpha }\left( \omega \right) $ is the power absorbed per unit
area. In Eq. (\ref{mwabsorption}) we have neglected a diamagnetic
contribution to the current response which becomes important at low
frequencies.

\begin{figure}[tbph]
\includegraphics[scale=1]{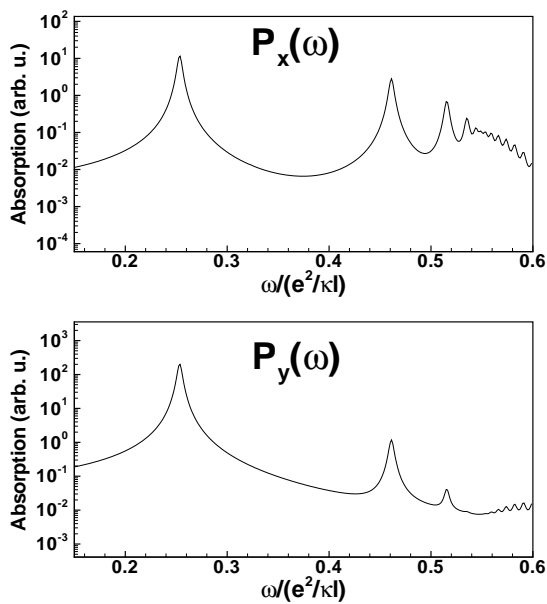}
\caption{Electromagnetic absorption $P_{\protect\alpha }\left( \protect%
\omega \right) $ in the helical phase for polarization in the $\widehat{%
\mathbf{x}}$ and $\widehat{\mathbf{y}}$ directions at bias $\Delta
_{B}/\left( e^{2}/\protect\kappa \ell \right) =1.0.$}
\label{absospirale}
\end{figure}

The absorption due to all but the gapless mode in the helical phase is shown
in Fig. \ref{absospirale}. Comparing with Fig. \ref{dispersionky}, we see
that all the other modes with the exception of the third mode (with
frequency near $0.3\left( e^{2}/\hslash \kappa \ell \right) $) are optically
active. The lowest gapped mode is the most intense one and its excitation is
strongly sensitive to the orientation of the polarization vector of the
electromagnetic wave. This is true at all bias voltages. The absorption
frequency does not change significantly with bias.

The absorption spectrum for the Skyrmion crystal phase is shown in Fig. \ref%
{absocristal}. In this case, all the modes (except the gapless phonon mode)
are equally active in the absorption and the absorption does not seem
sensitive to the polarization. There are thus qualitative differences
between the absorption in the helical and crystal phases that should help to
observe the transition between these two phases. Note that in the UP the
orbital pseudospin mode is gapless and does not lead to absorption.

\begin{figure}[tbph]
\includegraphics[scale=1]{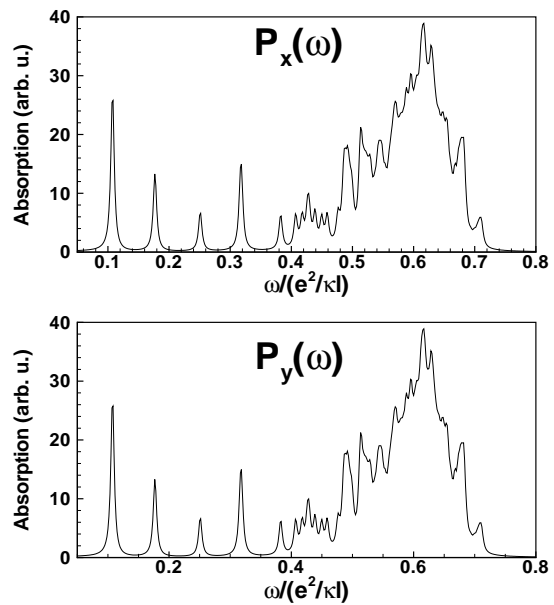}
\caption{Electromagnetic absorption $P_{\protect\alpha }\left( \protect%
\omega \right) $ in the crystal phase for polarisation in the $x$ and $y$
directions at bias $\Delta _{B}/\left( e^{2}/\protect\kappa \ell \right)
=0.2.$}
\label{absocristal}
\end{figure}

\section{DISCUSSION}

We have studied the phase diagram of the 2DEG in a graphene bilayer in the
Bernal stacking at filling factors $\nu =-1,3$ in Landau level $N=0$ when
orbital coherence is present in one of the layer. Our model uses a
tight-binding Hamiltonian that is simplified by working in the two-band
model. This simplification is justified since we are interested in the
low-energy excitations of the 2DEG. Moreover, we neglected the warping term,
an approximation that is valid at sufficiently strong magnetic field.
Finally, we assumed complete spin polarization. This last approximation may
fail $\nu =-1$ at finite bias when levels with spin up mix with levels with
spin down. A more exhaustive study\cite{Lambert} shows that, when this
approximation is not made, it is still possible to find states with orbital
polarization although at different filling factors and for different ranges
of bias than those studied in this paper.

The physics of the orbital coherent state is due to a competition between
the Coulomb exchange interaction and a Dzyaloshinskii-Moriya interaction
between the orbital pseudospins. This competition is responsible for a phase
diagram where the ground state evolves from a uniform state with
collectively oriented orbital pseudospins at small bias into a Skyrmion
crystal state and then into an helical state where the pseudospins rotate in
space. If the bias is further increased, the helical states transits into
the Skyrmion state again and then back to the uniform state. All three
states can be distinguished from their density of states and collective
excitations.

As was shown previously\cite{CoteOrbital}, the Goldstone mode due to
spontaneous orbital coherence in the uniform phase has the peculiarity of
being highly anisotropic. The dispersion is still highly anisotropic in the
helical state which is modulated in one direction only but is isotropic in
the Skyrmion crystal state. The three phases have one gapless mode and
several gapped modes. We have calculated that these latter modes lead to
absorption in the far-infrared region of the electromagnetic spectrum. In
the helical state, the absorption intensity is very sensitive to the
orientation of the polarization vector of the electromagnetic wave. This is
not the case in the Skyrmion crystal phase.

The helical and Skyrmion crystal phases each support a gapless phonon mode
which is accompanied by a motion of the pseudospins. We think that, in the
presence of disorder, these Goldstone modes should lead to strong absorption
of electromagnetic waves at very small frequencies. This calculation is
beyond the scope of this paper and we leave it for further work.\qquad

\begin{acknowledgments}
R. C\^{o}t\'{e} was supported by a grant from the Natural Sciences and
Engineering Research Council of Canada (NSERC) and J. P. Fouquet by a
scholarship from NSERC. Computer time was provided by the R\'{e}seau Qu\'{e}b%
\'{e}cois de Calcul Haute Performance (RQCHP).
\end{acknowledgments}

\end{document}